\documentclass{aa}
\usepackage{graphicx}
\usepackage{txfonts}
\usepackage{lipsum}
\usepackage{subcaption}             
\usepackage{lscape}                    
\usepackage{placeins}           
\usepackage[]{hyperref}
\hypersetup{
    citecolor=cyan,
    colorlinks=true,
    linkcolor=red,
    filecolor=red,      
    urlcolor=magenta,
    pdfpagemode=FullScreen,
    }
\usepackage{amsmath}    
\usepackage{color}     
\usepackage[normalem]{ulem}

\begin{document}

   \title{A pilot very long baseline interferometry study of the SQUAB quasar sample featuring multiple \textit{Gaia} detections}
   \subtitle{}

   \author{Y. Zhang\inst{1,2,3},
   T. An\inst{1,2},
   X. Ji\inst{1},
   Z. Zheng\inst{1},
   Y. Liu\inst{1,2},
   Q. Wu\inst{1},
   R. Lin\inst{1}
    \and
   S. Liao\inst{1,4}
          }

    \institute{Shanghai Astronomical Observatory, CAS, Nandan Road 80, Shanghai 200030, China\\
              \email{ykzhang@shao.ac.cn}
        \and
            State Key Laboratory of Radio Astronomy and Technology, A20 Datun Road, Chaoyang District, Beijing, P. R. China
        \and
             CSIRO Space and Astronomy, PO Box 1130, Bentley WA 6102, Australia
        \and School of Astronomy and Space Sciences, University of Chinese Academy of Sciences, No. 19A Yuquan Road, Beijing 100049, China
            }
   \date{Received September 30, 20XX}
   \authorrunning{Zhang et al.}

  \abstract
    {Our previous work identified a class of SDSS quasars with multiple \textit{Gaia} detections, classifying them as candidates for various astrophysical systems, such as quasar--star pairs, dual quasars, and gravitationally lensed quasars. In this paper we present a pilot very long baseline interferometry (VLBI) study targeting a radio-bright subsample and report the first high-resolution imaging results.}
    {By leveraging the milliarcsecond-scale resolution of VLBI and its precise astrometric coordination with \textit{Gaia}, we aim to refine the classification of these multiple matched sources, search for potential dual active galactic nuclei (AGNs), and assess the efficacy of the combined \textit{Gaia}-VLBI approach in resolving ambiguous quasar systems.} 
    {We cross-matched the Strange QUasar Candidates With ABnormal Astrometric Characteristics (SQUAB) quasar sample with the FIRST and NVSS catalogs, identifying 18 radio-emitting sources. The three brightest were selected for dual-frequency (1.6 and 4.9 GHz) VLBA observations. We performed VLBI imaging at both \textit{Gaia} positions, constructed spectral index maps, and estimated brightness temperatures to characterize the radio morphology and physical properties.}
    {For the three target sources, our VLBI observations reveal compact radio structures consistent with single AGNs at the primary \textit{Gaia} positions. No significant emission is detected at the secondary \textit{Gaia} locations. These results support the interpretation of the sources as quasar$-$star pairs, in line with earlier studies.}
    {
    This pilot study demonstrates the value of radio-VLBI high-resolution follow-ups on \textit{Gaia}-selected quasar systems with multiple counterparts, showing how they can unambiguously reveal the true nature of these systems and help remove contaminants from dual AGN candidate samples.
    }


   \maketitle

\section{Introduction} \label{sec1}
Quasars, as the most luminous class of active galactic nuclei (AGNs), lie at extreme distances and are compact in nature. 
They have traditionally been considered to exhibit negligible parallax and proper motions, making them ideal reference points for astrometric studies \citep[][]{1998RvMP...70.1393S,1998AJ....116..516M,2018A&A...616A..14G}. However, recent high-precision astrometric observations, particularly those from the \textit{Gaia} mission, have revealed that some quasars exhibit unexpected astrometric features. These anomalies include significant proper motions, parallaxes, or astrometric noise that are inconsistent with the traditional view of quasars as fixed points on the celestial sphere \citep[][]{2018A&A...616A...2L,2021A&A...649A...1G}. 
In a recent work, we identified a population of quasar candidates with anomalous astrometric properties by cross-referencing \textit{Gaia} Early Data Release 3 astrometric data with spectroscopic data from the Sloan Digital Sky Survey (SDSS), which we refer to as Strange QUasar Candidates With ABnormal Astrometric Characteristics \citep[SQUABs;][]{2022FrASS...922768W,2023MNRAS.524.1909J}. Within the SQUAB population, a particularly intriguing subset comprises SDSS quasars that are associated with multiple \textit{Gaia} counterparts within 1 arcsecond. We designate these as SQUAB sources with multiple \textit{Gaia} detections (hereafter SMGDs). There are several compelling questions regarding the physical nature and possible astrophysical origins of SMGDs.

While chance alignments of unrelated sources at different redshifts could account for some of the SMGD cases (e.g., quasar plus a foreground star or AGN), alternative interpretations include physically associated quasar pairs or gravitationally lensed quasars (LQs). Each of these scenarios carries distinct astrophysical implications and observational challenges, highlighting
the value of high-resolution and multiwavelength
follow-up studies of SMGDs.

Among the possible scenarios of SMGDs, quasar pairs (or dual AGNs) are of particular importance. 
Dual AGN systems are kiloparsec-separated pairs of supermassive black holes (SMBHs) brought together through dynamical friction and interactions with surrounding stars and gas \citep[e.g.,][]{2001ApJ...563...34M,2007Sci...316.1874M}. 
They represent the precursors of supermassive black hole binaries (SMBHBs), where the separations shrink to parsec scales and the two black holes eventually become gravitationally bound. 
Identifying dual AGN systems is therefore a crucial step toward tracing the merger-driven evolution of SMBHs.\citep{1980Natur.287..307B,2003ApJ...582..559V,2016ApJ...828...73K}.

According to hierarchical models of galaxy evolution, SMBHBs are a natural consequence of galaxy mergers. As galaxies collide and merge, their central SMBHs are expected to form a bound pair, eventually coalescing into a single, more massive black hole \citep{2002MNRAS.331..935Y,2017MNRAS.464.3131K}. This process is thought to be a key mechanism for the growth of SMBHs over cosmic time and is intimately linked to the coevolution of galaxies and their central black holes \citep[][]{2013ARA&A..51..511K}.

Despite their importance, confirming the existence of dual AGNs and SMBHBs, especially from subparsec to a few-kiloparsec scales, has proven challenging \citep[e.g.,][]{2018RaSc...53.1211A}. This difficulty arises from several factors. 
At cosmological distances, the angular separation of dual AGNs can be extremely small, often below the resolution limit of conventional optical telescopes \citep{2002MNRAS.331..935Y,2011MNRAS.410.2113B}.
While double-peaked emission lines in optical spectra have been proposed to be a signature of SMBHBs, they can also arise from other phenomena, such as outflows or rotating disks \citep[e.g.,][]{2012NewAR..56...74P,2010ApJ...716..866S}.
The orbital periods of SMBHBs can be very long, making it challenging to detect periodic variations that might indicate binary motion \citep{2015Natur.518...74G}.
There may also exist biases in the current detection methods that make certain types of dual AGNs easier to find than others \citep{2009ApJ...703L..86V}.
These challenges have led to a situation where, despite dedicated efforts, the number of confirmed dual AGNs and SMBHBs remains small. The majority of known dual AGN systems are either at low redshifts or have large physical separations ($>20$ kpc), with only a handful of dual quasars (DQs) with small separations found at high redshifts \citep{2020ApJ...897...86C}.

Very long baseline interferometry (VLBI) is increasingly recognized as a crucial tool for the study of dual AGNs and SMBHBs. Its unparalleled angular resolution on milliarcsecond scales makes it uniquely suited for directly resolving close pairs 
(see \citealt{2018RaSc...53.1211A} and the references therein). Over the years, VLBI observations have revealed a number of SMBH pairs with separations ranging from parsec to kiloparsec scales (see \citealt{2018ApJ...854..169L} for a review). 
Until now, the only known parsec-scale SMBHB system had been detected via VLBI imaging \citep[0402$+$379; see][]{2006ApJ...646...49R,2017ApJ...843...14B}. As for kiloparsec-scale dual AGNs, a few more cases were found serendipitously, including NGC~6240~\citep[$z=0.025$;][]{2004AJ....127..239G,2011AJ....142...17H,2022MNRAS.517.1791S}, NGC~5252~\citep[$z=0.023$;][]{2018MNRAS.480L..74M,2017MNRAS.464L..70Y}, J1502$+$1115~\citep[$z=0.39$;][]{2014Natur.511...57D,2014ApJ...792L...8W}, J0749$+$2255~\citep[$z=2.17$;][]{2023Natur.616...45C}, and J1536$+$0441~\citep[$z=0.389$;][]{2009ApJ...699L..22W,2010ApJ...714L.271B}. 
A promising dual AGN candidate was also studied with combined \textit{Hubble} Space Telescope and VLBI observations \citep[J1220$+$1126 at $z=1.9$;][]{2023ApJ...951L..18G}.

Many systematic VLBI-based searches for dual AGNs (or SMBHBs) failed to spot any multiple AGNs within a source. For example, \citealt{2011MNRAS.410.2113B} did not find any other flat-spectrum radio pairs from a sample of 3114 radio bright sources with multiband VLBI observations. \citealt{2011AJ....141..174T} matched 11 radio AGNs from 87 promising dual AGN (or SMBH) candidates that displayed double-peaked optical emission lines. The 11 radio sources were imaged with VLBI at 1.4 GHz, revealing only 2 sources with a single, weak component (17 mJy and 1.8 mJy). \citealt{2016MNRAS.459..820T} made another attempt based on 1279 VLBA Imaging and Polarimetry Survey (VIPS) sources \citep{2007ApJ...658..203H}, looking for double flat- or inverted-spectrum compact cores among a subsample of compact symmetric objects (CSOs) and CSO candidates, where no SMBHBs other than 0402+379 have been found.
These results highlight that, in addition to VLBI imaging capabilities, the selection of an oriented and intrinsically bright sample is critical when searching for dual AGNs or SMBHBs.

The search for SMBHBs and kiloparsec-scale dual AGNs with different separations is essential for several reasons.\ They allow us to test galaxy merger models and our understanding of SMBH growth, to probe the different stages of the SMBHB merger process, and to constrain the timescales of SMBHB evolution, which have implications for the "final parsec problem" in SMBHB mergers; they also inform predictions for gravitational wave emission from these systems  \citep{2003ApJ...596..860M}. 
The \textit{Gaia}-based multiple detection sample provides a bona fide and less biased parent sample that is well suited for VLBI follow-up studies.
By combining the high-precision astrometric measurements from \textit{Gaia} with the high-resolution imaging capabilities of VLBI, we have a unique opportunity to constrain the nature of these objects through a novel direct way.

In this paper we present the results of a pilot study that utilized this combined \textit{Gaia}-VLBI approach. We focus on a subset of SMGDs with strong radio emission, hereafter referred to as radio-detected SMGDs (R-SMGDs).
The target sources, namely J1433$+$4842, J1520$+$4211, and J0928$+$5707, were picked based on their radio flux densities determined via surveys such as the NRAO VLA Sky Survey (NVSS) and Faint Images of the Radio Sky at Twenty-centimeters (FIRST). 
Through VLBI observations of the three R-SMGDs, we obtained high-resolution radio images and accurate positions for comparison with \textit{Gaia} detections. The combination of radio morphology and spectral analysis enabled us to search for dual AGNs and place tighter constraints on the intrinsic properties of these objects. Moreover, this approach provides a unique opportunity to clarify the origins of \textit{Gaia}’s unusual astrometric measurements using the independent reference frame offered by VLBI.

We describe our sample selection based on the SMGD sample in Sect.~\ref{sec:sample}. 
In Sect.~\ref{sec:data} we present the data reduction and image analysis procedures. 
Section~\ref{sec:discussion} provides our identification of the target sources and discusses the prospects of the \textit{Gaia}--VLBI approach. 
Finally, Sect.~\ref{sec:conclusion} summarizes our main conclusions and outlines directions for future work.
Throughout this paper, we apply the $\Lambda$ cold dark matter cosmological model parameters with $H_{0}=70$~km~\,s$^{-1}$\, Mpc$^{-1}$, $\Omega_\mathrm{m}=0.3$, and $\Omega_{\Lambda}=0.7$.

\section{Sample selection}
\label{sec:sample}

The parent SMGD sample consists of 143 SDSS quasars. We extracted it from SQUAB sample A from our first SQUAB paper \citep[][]{2022FrASS...922768W} by excluding any single stars.
Our second SQUAB paper \citep[hereafter J2023]{2023MNRAS.524.1909J} concentrated on the examination of the SMGD sample; in that study, we conducted a series of optical identification methods to classify the sample.
A straightforward version of the procedure from J2023 is outlined below:
\begin{enumerate}
    \item   \textit{Gaia} detections of sources with obvious stellar features (e.g., proper motions or parallaxes greater than 5$\sigma$ and/or spectral and color characteristics) were marked as quasar--star pairs.     \item After filtering out the quasar--star pairs, we searched for optical spectral features, and sources showing double-peaked emission line features were marked as DQ candidates.
    \item A color-color analysis for the filtered source was conducted in parallel. Sources with colors similar to their \textit{Gaia} counterparts were marked as LQ candidates.
    \item Sources lacking adequate optical information (e.g., \textit{Gaia} non-detections in several bands) for the above identifications were marked as "REST."
\end{enumerate}
From the J2023 results, 65 quasar--star pairs, 2 DQ candidates, 56 LQs or LQ candidates, and 20 REST sources were compiled. 
However, using optical-only data for classification poses inherent limitations and biases.
This concern arises from the following facts:
(1) Confirmed sources remain scarce: only 13 LQs (out of the 56 LQ candidates) have been confirmed in the literature, and just 19 sources have high-resolution \textit{Hubble} Space Telescope imaging available for more reliable verification.
(2) Classification inconsistencies exist: for example, one reported DQ, J084129.77$+$482548.3 \citep[][]{2022NatAs...6.1185M}, is classified as a LQ candidate due to the lack of spectral confirmation. Two reported LQs, J081331.28$+$254503.0 \citep[][]{2015MNRAS.454..287J} and J234330.58$+$043557.9 \citep{2019arXiv191208977K}, are classified as quasar--star pairs due to \textit{Gaia}'s limitations. Two reported DQs and one reported LQ  continue to carry the REST label due to the lack of necessary features.

Given this context, in addition to the search for potential dual AGNs, we consider it essential to conduct high-resolution imaging of the SMGDs using the combined \textit{Gaia}-VLBI approach. This can provide evidence that we can use to robustly constrain their physical nature.
To identify radio-bright candidates suitable for VLBI observations within this sample, we performed a cross-match of the SMGDs with two major radio surveys: NVSS \citep[][]{1998AJ....115.1693C} and FIRST \citep[][]{1995ApJ...450..559B}.

The search method was as follows: First, we separately cross-matched the SDSS positions of the sample with NVSS and FIRST. Considering the different beam sizes and position errors of the surveys, different searching radii were used for the cross-matching: The searching radius for SDSS-NVSS was 10 arcsec and for SDSS-FIRST 3 arcsec. Second, we combined the two matched groups and generated a final sample consisting of their positional information, radio flux densities, and the separations between their \textit{Gaia} positions. 
Based on the cross-matching procedure, we built the R-SMGD sample, consisting of 18 radio sources with detectable radio emission in at least one of the surveys (see Table \ref{tab:big_table}; the classifications from J2023 are also listed in the last column).

As no prior VLBI studies have been reported for our R-SMGD sample (except for J1520$+$4211, which was included in sensitivity-limited astrometric surveys), we initiated new Very Long Baseline Array (VLBA) observations \citep[][]{2006ApJ...646...49R} to investigate their physical nature. The high angular resolution of the \textit{Gaia}$-$VLBI alignment allowed us to test their classifications with unprecedented precision and search for potential DQs.
To optimize the observing strategy and minimize the risk of VLBA non-detections, the brightest three sources in the table were proposed and approved for a pilot session, with follow-up observations on weaker sources planned for a later stage.

\section{VLBI observation and data analysis}\label{sec:data}
\subsection{VLBA observation and data reduction}\label{subsec:data1}
The VLBA observations of the three selected sources were conducted in 2023 February and March (project code: BZ94; see Table \ref{tab1:VLBAobs} for more information). The observing frequencies were 1.6 GHz (L band) and 4.9 GHz (C band).
The observations were performed using the digital downconverter system. Four 
128 MHz baseband channels on both left and right circular polarization and 2-bit sampling were utilized as the basic observing configuration, resulting in a total data rate of 4 Gbit s$^{-1}$ for the high-sensitivity VLBI observation.
Given that the target sources had not before been observed by VLBI, their high-resolution radio cores could be dimmer and more compact than their large-scale radio sources. Thus, the phase referencing technique was used for all three target sources.
The phase calibrators were searched for using the astrogeo Calibrator Search Engine\footnote{https://astrogeo.org/calib/search.html}. The phase calibrators used for the three sources were all standard calibrators (with a "C" flag) and were within 2 degrees of the target source position. This guaranteed the accuracy and imaging quality of the phase that applied to the target source. We used a 5-minute phase-referencing cycle time (4 min on target and 1 min on calibrator) for each target source in both observing bands. Since the two \textit{Gaia} positions of each source are not very distant ($\sim$ 1 arcsec), we only used the first \textit{Gaia} position (J2000) as the target position, i.e., $\mathrm{RA}=14^\mathrm{h}33^\mathrm{m}33\fs0316$, $\mathrm{Dec}=+48\degr42\arcmin27\farcs7752$ for J1433$+$4842; $\mathrm{RA}=15^\mathrm{h}20^\mathrm{m}39\fs7218$, $\mathrm{Dec}=+42\degr11\arcmin11\farcs4524$ for J1520$+$4211; and $\mathrm{RA}=09^\mathrm{h}28^\mathrm{m}53\fs5310$, $\mathrm{Dec}=57\degr07\arcmin35\farcs9477$ for J0928$+$5707.

After the observations, the raw data from each antenna were transferred to the VLBA correlation center in Socorro (New Mexico, USA) for correlation at the Distributed FX (DiFX) correlator \citep[][]{2011PASP..123..275D} with a 2s integration time. For the second \textit{Gaia} position of each source, we also requested the additional correlation of the same data using the second \textit{Gaia} position as the target phase center, which helps reduce the smearing effect and increases the imaging quality for both \textit{Gaia} positions. We then downloaded the correlated data and conducted additional data calibration and imaging using the VLBI pipeline at the China Square Kilometre Array (SKA) Regional Centre \citep[][]{2019NatAs...3.1030A,2022arXiv220613022A}. The pipeline takes advantage of the US National Radio Astronomy Observatory (NRAO) \texttt{AIPS} (Astronomical Image Processing System) software \citep[][]{2003ASSL..285..109G} and the \texttt{ParselTongue} tool to translate \textsc{\texttt{AIPS}} tasks into \texttt{Python} scripts \citep{2006ASPC..351..497K}  and automatically calibrates the data with minor human intervention. The calibration steps follow the standard calibration procedure of the \texttt{AIPS} Cookbook\footnote{see Appendix C in http://www.aips.nrao.edu/cook.html} and can be found in \cite{2022ApJ...937...19Z} and \cite{2023SSPMA..53v9507L}.
After calibration, the calibrated visibility data were exported by the pipeline and loaded using the Caltech \texttt{DIFMAP} software package \citep[][]{1997ASPC..125...77S} for imaging. We made use of the hybrid mapping procedure, which includes several loops of CLEAN and self-calibration steps in \texttt{DIFMAP}. After that, iterations of phase and amplitude self-calibration with decreasing integration time (divided by 2 in each iteration, down to 2 min) were conducted to generate the final images. Note that for source J1433$+$4842, only phase self-calibration was performed, to avoid spurious detections due to the weakness of the source \citep[e.g.,][]{2008A&A...480..289M}. 
We performed this calibration procedure for both the first and second \textit{Gaia} positions for each source. This was accomplished by separately processing the data whose correlation phase center corresponded to the respective \textit{Gaia} position, to eliminate the bandwidth smearing effect, which would otherwise blur the detection of sources far from the imaging phase center.
For source J1433$+$4842, we also did an extra shift-and-CLEAN procedure for a newly found extending component, which lies $\sim$ 280 mas from its first \textit{Gaia} position (see Sect. \ref{subsec:data2} for more details). This extra step was done in \texttt{AIPS} with the task \texttt{UVFIX} to shift the imaging phase center to the component peak before any channel or time averaging of the data was conducted. Then the shifted visibility data were exported again for another imaging round in \texttt{DIFMAP} in order to get a better image of the extending component.
The image pixel cell sizes were adopted following the recommendations of \texttt{DIFMAP} during the CLEAN process: 1.8 $\times$ 1.8 mas at 1.6 GHz and 0.5 $\times$ 0.5 mas at 5 GHz.

Following the imaging procedure, we performed model-fitting in \texttt{DIFMAP} on the self-calibrated visibility data to parameterize the VLBI components.
During the fitting process, we initially adopted a circular Gaussian brightness distribution model centered on the component peaks of each detected source feature. 
If the residual peak remained beam-like in shape, an additional circular model was applied. 
In cases where the residual appeared too compact or exhibited an irregular structure, a point-source (delta-function) model was applied instead. 
The fitting process was iterated until no significant residuals above the $5\sigma$ level remained within the CLEANed area.
Parameter descriptions from the fitting results are presented in Sect. \ref{subsec:data2}.
The uncertainties in the fitted component parameters were primarily estimated following the method described by \citealt{1999ASPC..180..301F}. For flux density errors, we accounted for an additional 5\% uncertainty introduced during the calibration process. In terms of VLBI positional uncertainties, we adopted the widely used $\frac{\theta_{beam}}{2S/N}$ formula (where $\theta_{beam}$ is the full width at half maximum size of the beam and S/N is the signal-to-noise ratio of the component) as the measurement error for VLBI observations \citep[e.g.,][]{2010MNRAS.409L..64Y,2020NatCo..11..143A}. Moreover, since we used phase-referencing, positional errors can also arise from the calibrator's position. As a result, the uncertainties in the phase calibrator's position are considered the dominant source of systematic error. The final positional uncertainties are thus the quadratic means of these two errors.

\begin{table*}
\caption{Information about the VLBA observations.}
\label{tab1:VLBAobs}
\centering
\begin{tabular}{ccccccc}
\hline
Code  &  Frequency & Date & Target & Calibrator & Time & Telescopes$^{\rm a}$ \\
  & (GHz) & (YYYY/MM/DD) & (J2000) & (J2000) & (h) & \\
\hline
bz94a1&1.63&2023/02/19&J1433$+$4842&J1439$+$4958&3.5&BR,FD,HN,KP,LA,NL,OV,PT,SC \\
bz94a2&1.63&2023/03/10&J1520$+$4211&J1521$+$4336&3.5&BR,FD,KP,LA,NL,OV,PT,SC \\
bz94a3&1.63&2023/02/13&J0928$+$5707&J0927$+$5717&3.5&BR,FD,KP,LA,NL,MK,OV,PT,SC \\
bz94b1&4.87&2023/02/27&J1433$+$4842&J1439$+$4958&3&BR,HN,LA,NL,OV,PT,SC \\
bz94b2&4.87&2023/03/08&J1520$+$4211&J1521$+$4336&3&BR,FD,HN,KP,LA,NL,OV,PT,SC \\
bz94b3&4.87&2023/02/14&J0928$+$5707&J0927$+$5717&3&BR,FD,HN,KP,LA,MK,NL,OV,PT,SC \\
bz94b4$^{\rm b}$&4.87&2023/03/01&J1433$+$4842&J1439$+$4958&3&BR,FD,HN,KP,LA,NL,OV,PT,SC \\
\hline
\end{tabular}

\tablefoot{Column 1: Observation code of the VLBA sessions. Column 2: Observing frequency. Column 3: Observation date. Column 4: The target of the observation session. Column 5: Phase calibrator of the phase referencing session. Column 6: Overall observing time of the session. Column 7: Participating telescopes of the VLBA observation.
\\
$^{\rm a}$VLBA telescopes participating in the observations: BR (Brewster), FD (Fort Davis), HN (Hancock), KP (Kitt Peak), LA (Los Alamos), MK (Mauna Kea), NL (North Liberty), OV (Owens Valley), PT (Pie Town), and SC (Saint Croix). \\ 
$^{\rm b}$ Since too many antennas failed in bz94b1, bz94b4 was performed as a compensatory session for J1433 in the C band.}
\end{table*}

\begin{table*}
\caption{CLEAN image results for each target source.}
\label{tab2:imparms}
\centering
\begin{tabular}{cccccccc}
\hline
Target & Date & Freq. & Peak & $\sigma$ & Bmaj & Bmin & Bpa \\
(J2000) & & (GHz) & (mJy/b) & (mJy/b) & (mas) & (mas) & (mas) \\
\hline
J1433$+$4842 & 2023/02/19 & 1.63  & 1.81  & 0.025 & 12.0  & 6.8  & 2.1  \\
... & 2023/03/01 & 4.87  & 0.86  & 0.019 & 5.0  & 2.3  & $-$9.4  \\
J1520$+$4211 & 2023/03/10 & 1.63  & 32.50  & 0.029 & 14.9  & 7.2  & 17.8  \\
... & 2023/03/08 & 4.87  & 25.96  & 0.013 & 4.1  & 1.9  & 10.8  \\
J0928$+$5707 & 2023/02/13 & 1.63  & 3.97  & 0.024 & 12.9  & 5.2  & 5.3  \\
... & 2023/02/14 & 4.87  & 1.99  & 0.018 & 3.9  & 1.4  & 5.1  \\
\hline
\end{tabular}
\tablefoot{Column 1: Target name. Column 2: Observing date in the format of YYYY/MM/DD. Column 3: Observation frequency. Column 4: The peak intensity of the image. Column 5: The post-CLEAN rms noise of the image. Columns 6 -- 8: The FWHM major axis, minor axis, and the position angle (from north to east) of the synthesized elliptical beam.}
\end{table*}

\subsection{Image analysis} \label{subsec:data2}
Our VLBI observations of the three target sources (J1433$+$4842, J1520$+$4211, and J0928$+$5707) yielded high-resolution images at both 1.6 GHz and 4.9 GHz.
For each source, we present the CLEAN wide-field image at 1.6~GHz, covering both \textit{Gaia} positions, along with the CLEAN zoomed-in images of individual components at 1.6 and 5~GHz, with astrometric positions clearly marked.
Appendix \ref{app:wideim} displays the large-scale images, where square boxes highlight the \textit{Gaia} positions or components. The corresponding high-resolution zoom-ins are shown in Appendix \ref{app:smallim}. In the various figures, we label the first and second \textit{Gaia} detections as GAIA1 and GAIA2, respectively.
The CLEAN image results are listed in Table \ref{tab2:imparms}. 

For J1433$+$4842, from Figs. \ref{fig:J1433_big} and \ref{fig:J1433_small}, we detected a compact radio component close to the GAIA1 position, while no significant radio emission is detected at its GAIA2 position.
We also discovered an extended jet-like structure (hereafter EXT1) between the two \textit{Gaia} positions, at both 1.6 and 5 GHz. The peak intensity of EXT1 is located at $\mathrm{RA}=14^{\mathrm{h}} 33^{\mathrm{m}} 32\fs9876$ and $\mathrm{Dec}=48\degr 42\arcmin 27\farcs5872$, approximately 280 mas southwest of GAIA1. 
For sources J1520$+$4211 and J0928$+$5707, the imaging results are similar: only GAIA1 is detected, with no significant radio emission at the GAIA2 positions. At both 1.6 and 5 GHz, GAIA1 in J1520$+$4211 exhibits a core-jet morphology, while J0928$+$5707 appears as a compact source.

\subsubsection{Flux density measurements}

From the imaging and model-fitting procedures described above (see Sect. \ref{subsec:data1}), we obtained high-resolution positions and flux density measurements of each source component.
For each source, the VLBI core component associated with GAIA1 is well described by a single circular Gaussian model, which was used to determine the VLBI position (Table~\ref{tab3:subimparms}) and to calculate the brightness temperature (Table~\ref{tab4:tb}).

In Table \ref{tab3:subimparms}, for each marked feature in Col. 2, Cols.~3--5 report the optical positions and errors from \textit{Gaia} detection. Columns~6--11 report the imaging and model fitting results from our VLBI observation; Col.~7 represents the peak flux densities of each feature. For non-detections, the 5$\sigma$ upper limits are presented. Columns~8 and 9 are the positional offsets between the \textit{Gaia} detection and the VLBI core component position along RA and Dec, separately. Column~10 lists the local post-CLEAN rms noise derived from the pixels within the zoomed-in area (shown in Appendix \ref{app:smallim}). Column~11 lists the total flux density of each radio feature.
This was obtained by summing the fitted core component and the remaining model components associated with the feature derived during the model fitting process.

\begin{table*}
\caption{Flux density and brightness temperature of the core component.}
\label{tab4:tb}
\centering
\begin{tabular}{ccccccccc}
\hline
Source      & ${\rm \nu}$& F$_{\rm \nu, core}$& D$_{\rm core}$& T$_{B}$ \\
(J2000)     & (GHz)      & (mJy)              & (mas)         & (K) \\ \hline
J1433$+$4842& 1.63       & 1.19$\pm$0.08        & $\le$1.55     & $\ge$5.3$\times$10$^{8}$\\
            & 4.87       & 0.92$\pm$0.05        & 0.63$\pm$0.07 & (2.8$\pm$0.5)$\times$10$^{8}$\\
J1520$+$4211& 1.63       & 31.60$\pm$1.58       & 0.59$\pm$0.01 & (6.2$\pm$0.3)$\times$10$^{10}$\\
            & 4.87       & 25.00$\pm$1.25       & $\le$0.06     & $\ge$5.6$\times$10$^{11}$\\
J0928$+$5707& 1.63       & 4.09$\pm$0.21        & 1.22$\pm$0.05 & (3.4$\pm$0.3)$\times$10$^{9}$\\
            & 4.87       & 1.03$\pm$0.06        & $\le$0.21     & $\ge$3.2$\times$10$^{9}$\\
\hline
\end{tabular}
\\
\tablefoot{Column 1: Target name. Column 2: Frequency. Column 3: The flux density of the fitted core component in each source's GAIA1. Column 4: The fitted FWHM size of the core component. Column 5: Brightness temperature.}
\end{table*}

Table \ref{tab4:tb} presents the core parameters associated with GAIA1. Column 3 denotes the integrated flux densities of the fitted core components. The full width at half maximum size of the core is listed in Col. 4, where the size upper limits for unresolved components
are derived based on \citealt{2005AJ....130.2473K} (i.e., if the fitted component size is smaller than this value, this value is used as the upper limit for the component size).
The brightness temperatures of the core component in Col. 5 were calculated as \citep[e.g.,][]{1982ApJ...252..102C}
\begin{equation}
T_{\mathrm B}=1.22 \times 10^{12} (1+z) \frac{S_{\nu}}{\theta_\mathrm{comp}^{2}\nu^{2}} \, {\mathrm K},
\end{equation}
where $S_{\nu}$ is the component flux density (in Jy),  $\nu$  the observing frequency of the image (in GHz), and ${\theta_\mathrm{comp}}$ is the full width at half maximum of the circular Gaussian component diameter (in mas). If ${\theta_\mathrm{comp}}$ is an upper limit, the calculated $T_{\mathrm B}$ was used as a lower limit. 
The fact that the  $T_{\mathrm B}$  values of all the core components exceed $10^{8}$K indicates that the GAIA1 features in our sample are predominantly powered by AGN activity.

\subsubsection{Spectral index map} \label{sec:spx}
To gain further insights into the physical nature of these sources, we generated spectral index maps between 1.6 GHz (L band) and 4.9 GHz (C band). To avoid ambiguity, we defined the radio spectral index ($\alpha$) according to the power-law relation S$_{\nu} \propto \nu^{\alpha}$, where S$_{\nu}$ is the flux density at frequency ${\nu}$. In this convention, a negative $\alpha$ value indicates that the flux density decreases with increasing frequency.

The phase-referencing observations provide high-quality positions of the radio source at the milliarcsecond level, making the alignment of maps from different frequencies convenient and accurate. The positions in Table \ref{tab3:subimparms} were used as image centers for the alignments.
To account for differences in beam size and pixel scale between images at different frequencies, we adopted the mapping parameters (i.e., restoring beam size, pixel size, and map size) from the lower-frequency L-band images as a reference. The higher-frequency C-band images were then reconstructed using the same beam and pixel configurations as the referenced ones to ensure consistency in the mapping of spectral index values.
Before computing the spectral index maps, we applied a masking procedure to exclude pixels with intensities below 3$\sigma$ of the local image rms noise. This step helped eliminate spurious $\alpha$ values arising from noise-dominated regions.

\begin{figure*}
\centering
\includegraphics[width=0.35\linewidth]{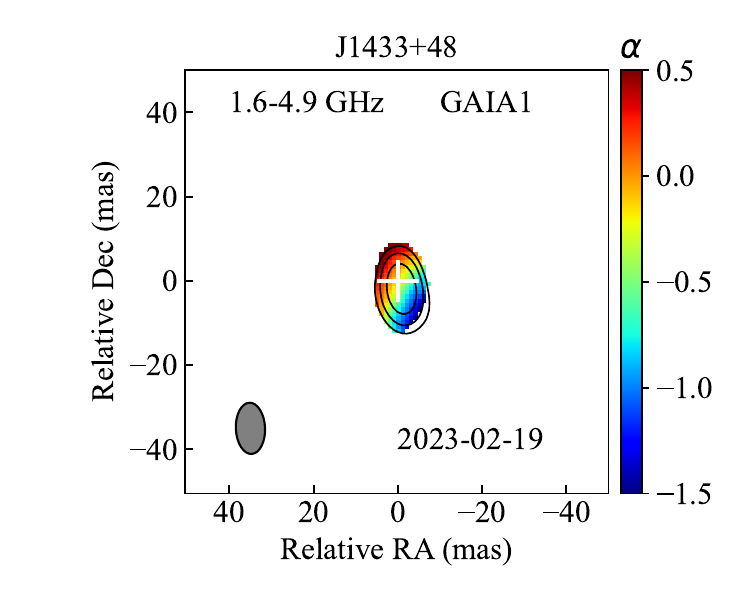} 
\includegraphics[width=0.35\linewidth]{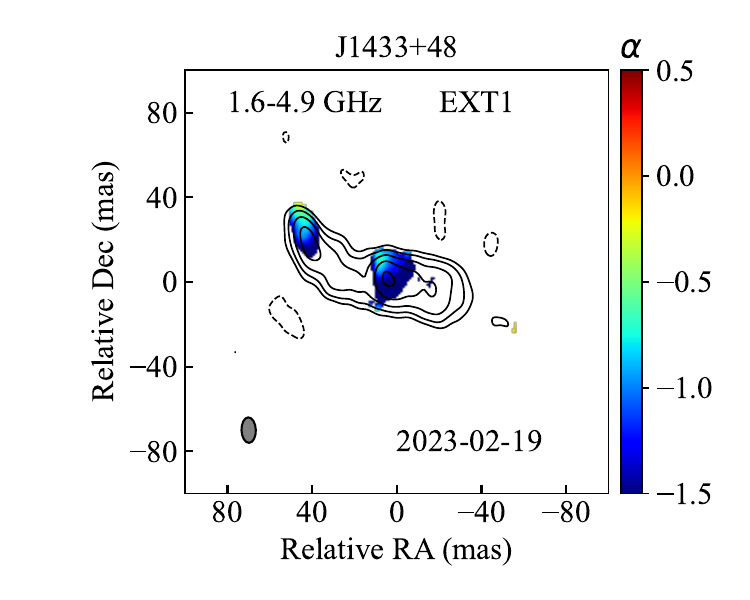} \\
\includegraphics[width=0.35\linewidth]{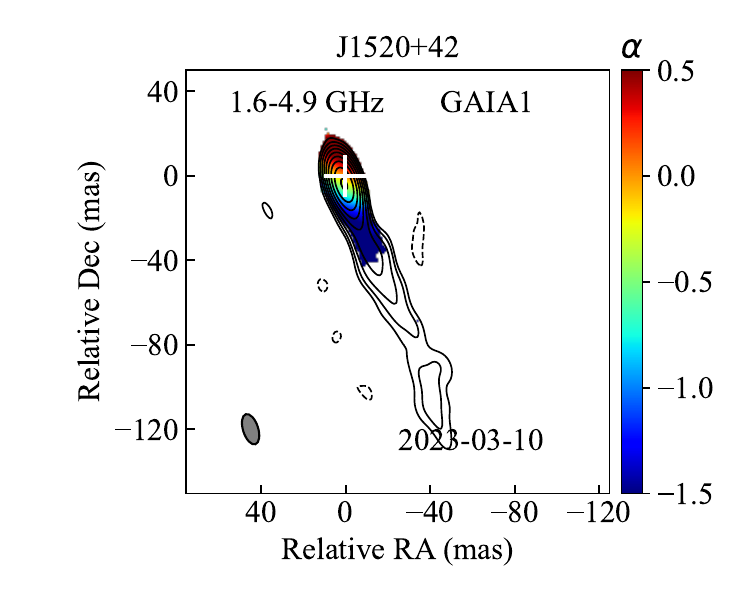} 
\includegraphics[width=0.35\linewidth]{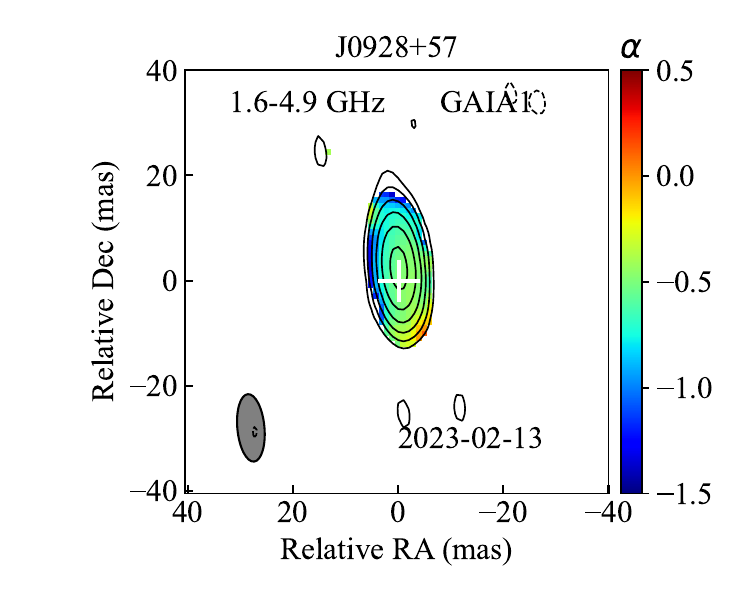} 
\caption{Spectral index maps of the target sources.
Contours denote the low-frequency (1.6 GHz) intensity map, with the corresponding synthesized beam shown in the bottom-left corners. 
The lowest contour corresponds to $4\sigma$, and subsequent levels increase by factors of two. 
The color scale represents the spectral index distribution between 1.6 and 4.9-GHz (see Sect.~\ref{sec:spx}).
\label{fig:spix}}
\end{figure*}

Following the alignment and mapping procedures described above, spectral index maps for each radio feature were successfully generated, as shown in Fig. \ref{fig:spix}. To assess the reliability of these maps, we also present the corresponding spectral index error maps (see Fig. \ref{fig:spixerr}), computed through numerical error propagation following standard methods \citep[e.g.,][]{2024ApJ...971...39R}.

Figure \ref{fig:spix} (top panels) displays the spectral index maps for both the GAIA1 and EXT1 of J1433$+$4842. The radio core in GAIA1 shows a flat-spectrum feature ($\alpha \approx 0.0$).
The spectral index map of EXT1 reveals a predominantly steep-spectrum jet structure with $\alpha \le -0.5$ across most of its extent; it becomes steeper toward the western end.
The spectral index map for J1520$+$4211 (Fig. \ref{fig:spix}, bottom-left panel) shows a typical core-jet structure. The core region exhibits a flat spectrum with $\alpha \approx -0.2$, typical of synchrotron self-absorbed emission from the jet base. Toward the southwest, the spectrum steepens progressively, reaching $\alpha \approx -1.0$ at the farthest extent of the detected emission. 
For source J0928$+$5707 (Fig.\ref{fig:spix}, bottom-right panel), a flat-spectrum core with $\alpha \le -0.5$ is revealed.

\section{Discussion}\label{sec:discussion}

\subsection{Source classification from VLBI identification}
Based on our VLBI imaging results for the three target sources (J1433$+$4842, J1520$+$4211, and J0928$+$5707), the observed radio cores are in close alignment with their first \textit{Gaia} detections (GAIA1). The compactness, high brightness temperature (e.g., $\ge 10^{8}$ K), and flat-spectrum characteristics of the radio cores strongly indicate a AGN nature for their GAIA1 features. Although no significant radio emission at their GAIA2 positions was detected, our VLBA observations provide a 5$\sigma$ upper flux density limit of approximately 0.1 mJy/b. 
Based on the \textit{Gaia} G-band detections, the optical flux ratios, F$_{\rm GAIA1,opt}$/F$_{\rm GAIA2,opt}$, for J1433$+$4842, J1520$+$4211, and J0928$+$5707 are 5.6, 1.0, and 0.3, respectively. 
On the basis of strong lensing laws, we would expect comparable flux ratios in the radio band. 
This corresponds to predicted radio peak intensities at 1.6 GHz for the GAIA2 counterparts of $\sim$ 0.2, 31.4, and 11.6 mJy\,beam$^{-1}$, respectively. 
These estimated values exceed our derived flux density thresholds for GAIA2, thereby allowing us to rule out the LQ scenario.
Given that no additional compact radio components are observable aside from their GAIA1 features, the most plausible explanation for their optical GAIA2 detections is a foreground star. 

\subsection{Information on individual sources}
\subsubsection{J1433$+$4842}
(

\citealt{2023MNRAS.524L..38W} first studied J1433$+$4842 using Enhanced Multi Element Radio-Linked Interferometer Network (eMERLIN) observations, focusing on its astrometric jitter detected by \textit{Gaia}. Their study, which used a selection method based on \textit{Gaia}'s astrometric excess noise, detected a single compact radio source close to the \textit{Gaia} position with a small \textit{Gaia}-radio offset. They suggested that the observed astrometric jitter could be due to contamination from nearby sources rather than being indicative of a dual AGN system.
\citealt{2024ApJ...961..233S} (hereafter S24) included J1433$+$4842 in their VaDAR (Varstrometry for Dual AGN using Radio interferometry) pilot study. Based on the jet morphologies and spectral index in S-band (2-4 GHz) and X-band (8-12 GHz) Very Large Array (VLA) observations, they identified J1433$+$4842 as a single jetted AGN rather than a dual AGN candidate.

Our high-resolution VLBI results for J1433$+$4842 are consistent with previous identifications and additionally reveal a distant jet feature, EXT1, that originated from the radio cores in GAIA1. By superimposing our radio features onto the VLA images from S24, we observe a strong concordance between the VLA images and our VLBI features of GAIA1 and EXT1 (see Fig. \ref{fig:target1_vla}).

\subsubsection{J1520$+$4211}
Our VLBI images of J1520$+$4211 reveal a core-jet morphology, which is characteristic of radio-loud AGNs. At the GAIA1 position, the radio core exhibits peak intensities of 32.50 $\pm$ 1.63 mJy/beam at 1.6 GHz and 25.96 $\pm$ 1.30 mJy/beam at 4.9 GHz. The total flux densities are measured to be 42.37 $\pm$ 2.12 mJy at 1.6 GHz and 28.90 $\pm$ 1.45 mJy at 4.9 GHz.
At the GAIA2 position, no compact radio emission was detected. We derive 5$\sigma$ upper limits of 0.11 mJy/beam at 1.6 GHz and 0.07 mJy/beam at 4.9 GHz.

\subsubsection{J0928$+$5707}
We identified a single, compact radio core with flux densities of 4.09 $\pm$ 0.21 mJy at 1.6 GHz and 1.03 $\pm$ 0.06 mJy at 4.9 GHz for J0928$+$5707 at the GAIA1 position. We established 5$\sigma$ upper limits of 0.12 mJy/beam at 1.6 GHz and 0.09 mJy/beam at 4.9 GHz for its GAIA2 location.

\subsection{Prospects of the Gaia-VLBI method}
\citealt{2024ApJ...969...36X} (hereafter Xu2024) performed a VLBI imaging study of four dual-AGN systems that had been confirmed by \citealt{2015ApJ...799...72F}, both spatially and spectroscopically. The four dual AGNs show well-separated, spatially correlated double morphology in their VLA A-configuration 5 GHz observations \citep[see][]{2015ApJ...815L...6F}, with each component showing steep spectra ($\alpha \le -0.5$).
In the Xu2024 results, none of them show detection (above 5$\sigma$) on both components down to a imaging sensitivity of $\sim$ 0.025 mJy/beam. Two of them are revealed to have a single VLBI component at one of the component centers, with flux densities of 0.2 and 0.4 mJy, respectively. The VLA-detected radio components suggest that radio emission from these dual AGNs are dominated by less energetic jets, AGN-driven outflows, corona and star formation.

Based on the "standard routine" of directly looking for dual AGNs or SMBHBs in the literature \citep[see, e.g.,][and our Sect. \ref{sec1}]{2018JApA...39....8R,2018BSRSL..87..299D}, discoveries of dual AGNs and SMBHBs heavily rely on optical imaging and spectroscopic data to highlight the spatial correlation between two objects. Searching for dual AGNs or binary systems purely through radio observations is still challenging, partly due to the current fact that most quasars in pair systems tend to be radio-quiet or lobe-dominated galaxies with weak radio cores.

Very long baseline interferometry  plays a crucial role in identifying active radio cores in AGN systems and is indispensable for searching for dual AGNs at higher redshifts.
For instance, at redshifts ($z$) $>0.5$, a projected physical separation of 1 kpc corresponds to an angular separation of less than 0.2 arcsec, a scale at which only VLBI can directly resolve the two components in radio observations.
In recent years, different approaches utilizing \textit{Gaia} data have proven to be another prominent way of looking for dual AGNs, especially at high redshifts (i.e., $z>1$); they are called the "varstrometry" method \citep[e.g.,][]{2020ApJ...888...73H,2023ApJ...958...29C}  and the "\textit{Gaia} multi-peak" (GMP) method \citep[see][]{2022NatAs...6.1185M,2023A&A...680A..53M}.
In the varstrometry method, sub-kiloparsec AGN pairs are searched for by looking at the variations in the photocenter (or "jittering") from \textit{Gaia} astrometric parameters (e.g., ${astrometric\_excess\_noise}$) of unresolved sources; this method has been successfully used to discover a few candidates and confirmed sources \citep[e.g.,][]{2021NatAs...5..569S,2023MNRAS.524L..38W}.
The GMP method takes advantage of the ${ipd\_frac\_multi\_peak}$ parameter from the \textit{Gaia} sources. This method can detect potentially multiple sources within a \textit{Gaia} point spread function and hopefully help identify DQs or LQs at source separations in the range 0.1" to 0.7" \citep[see][]{2022NatAs...6.1185M}.

Our approach of combining VLBI and high-resolution  \textit{Gaia} observations complements existing methods for searching for dual AGNs, particularly in excluding contaminants among dual AGN candidates with ambiguous optical features.
Thanks to the milliarcsecond-scale angular resolution achievable through this method, dual AGNs  can be directly resolved  with projected separations down to several parsecs at moderate redshifts (assuming a resolution of $\sim$ 1 mas for redshifts above 0.5).
These results motivate our ongoing VLBI follow-up studies of the remaining fainter R-SMGDs (see Table~\ref{tab:big_table}).

\begin{table*}[htbp]
\caption{VLBI characteristics and diagnostics of the target sources reported in this paper.}
\label{tab:source_sum}
\centering
\begin{tabular}{cccll}
\hline
Source      & z & Radio Morphology & Spectral index & VLBI \textit{Gaia} correlation \\
\hline
J1433$+$4842& 1.75       & Core + extended jet        & Core: $\alpha \approx 0$     & GAIA1: core               \\
            &            &                            & Jet: $\alpha\le -0.5$     & GAIA2: not detected \\ \hline
J1520$+$4211& 0.49       & Core-jet                   & Core: $\alpha \approx -0.2$  & GAIA1: core               \\
            &            &                            &                               & GAIA2: not detected    \\ \hline
J0928$+$5707& 1.68       & Core-jet                   & Core: $\alpha \ge -0.5$      & GAIA1: core               \\
            &            &                            &                               & GAIA2: not detected   \\ \hline
\hline
\end{tabular}
\end{table*}

\section{Conclusion} \label{sec:conclusion}
In this study we conducted a dual-frequency VLBI imaging study of the three radio-brightest sources in our compiled R-SMGD sample: J1433$+$4842, J1520$+$4211, and J0928$+$5707. We have provided the first high-resolution images of these sources.
The present work serves as a pilot investigation to evaluate the effectiveness of the SQUAB classification from J2023 and to search for potential dual AGN systems through a combined \textit{Gaia}-VLBI approach.

For each target, we obtained VLBI images at 1.6 GHz and 4.9 GHz for both \textit{Gaia} components. We parameterized the source radio components to estimate key physical properties (e.g., brightness temperature).\ We produced spectral index maps to help constrain the nature of the radio emission and shed light on the origin of their multiple \textit{Gaia} detections. A summary of our observational results is provided in Table \ref{tab:source_sum}.

All three sources show clear evidence of AGN activity coinciding with their primary \textit{Gaia} detections (GAIA1), while no significant radio emission was detected at the positions of their secondary \textit{Gaia} components (GAIA2). In the case of J1433$+$4842, we also detected extended radio emission spatially between the two \textit{Gaia} positions, which is interpreted as jet structure originating from the AGN at GAIA1.
Based on these results, we confirm that all three sources are best explained as quasar--star pairs, in agreement with previous studies. 

Our pilot study on the R-SMGDs proved the validity of the SQUAB criteria and demonstrates that the incorporation of \textit{Gaia} and VLBI can be a promising method for searching for dual AGNs and for ruling out contaminants among dual AGN candidates.
With the sensitivity achieved in our pilot observations, we are able to 
detect radio emission down to $\sim$ 0.1 mJy beam$^{-1}$, recovering flux densities as low as 10\% of the VLA total flux densities for fainter sources.
Encouraged by these results, our follow-up studies (Zhang et al. in prep.) on the remaining R-SMGDs --
including several DQ and LQ candidates identified by J2023 -- are expected to expand the census of dual AGN candidates.
Future studies that use more sensitive optical facilities (e.g., JWST) and upcoming \textit{Gaia} data releases will likely yield additional valuable candidates for expanding \textit{Gaia}-VLBI investigations into DQ searches.

\begin{acknowledgements}
We are deeply grateful to the anonymous referee for the profound feedback, which has greatly enhanced both the scientific merit and the clarity of this work.
This work is supported by
the Strategic Priority Research Program of the Chinese Academy of Sciences, Grant No. XDA0350205.
This work used resources of China SKA Regional Center prototype funded by the Ministry of Science and Technology of the People’s Republic of China and the Chinese Academy of Sciences \citep{2019NatAs...3.1030A,2022SCPMA..6529501A}.
YZ is supported by
the National SKA Program of China (grant no. 2022SKA0120102),
the China Scholarship Council (No. 202104910165),
the Shanghai Sailing Program under grant number 22YF1456100.
YZ thanks the warm host and the helpful comments from Ivy Wong in CSIRO Space\&Astronomy, Australia. 
YZ thanks the valuable suggestions and comments from Sándor Frey and Krisztina \'Eva Gab\'anyi in Konkoly Observatory, Hungary.
TA acknowledge the support from the Xinjiang Tianchi Talent Program.
This work has made use of data from the European Space Agency (ESA) mission \textit{Gaia} (\url{https://www.cosmos.esa.int/gaia}), processed by the \textit{Gaia} Data Processing and Analysis Consortium (DPAC, \url{https://www.cosmos.esa.int/web/gaia/dpac/consortium}).
This work made use of public data from the data archive of the National Radio Astronomy Observatory (NRAO; \url{https://data.nrao.edu/}). The National Radio Astronomy Observatory is a facility of the National Science Foundation operated under a cooperative agreement by Associated Universities, Inc.
\end{acknowledgements}

\bibliographystyle{aa} 
\bibliography{refs} 

\begin{appendix}

\begin{sidewaystable*}
\section{Additional tables}
\caption{VLBI results of each source feature in this paper.}
\label{tab3:subimparms}
\centering
\begin{tabular}{ccccccccccc}
\hline
Source & Mark& Position& $\Delta_{ra}$& $\Delta_{dec}$& Freq.& Peak   & Offset$_{\rm RA}$& Offset$_{\rm Dec}$& $\sigma$   & S$_{\rm int}$ \\
(J2000)&     & (J2000) & (mas)        & (mas)         & (GHz)& (mJy/b)& (mas)            & (mas)             & ($\mu$Jy/b)& (mJy) \\
(1)    & (2) & (3)     & (4)          & (5)           & (6)  & (7)    & (8)              & (9)               & (10)       & (11) \\
\hline
J1433$+$4842& GAIA1& 14: 33: 33.0316+48: 42: 27.7752& 0.3& 0.3& 1.63& 1.16$\pm$0.07 & $-$0.8$\pm$0.2& $-$1.8$\pm$0.2& 38& 1.19$\pm$0.08 \\
...         & ...  & ...                            & ...& ...& 4.87& 0.86$\pm$0.05 & 0.5$\pm$0.2   & 0.5$\pm$0.1   & 17& 0.92$\pm$0.05 \\
...         & GAIA2& 14: 33: 32.9630+48: 42: 27.4198& 4.9& 4.2& 1.63& $\le$0.11    & $/$           & $/$           & 21& $/$ \\
...         & ...  & ...                            & ...& ...& 4.87& $\le$0.09    & ...           & ...           & 17& $/$ \\
...         & EXT1 & 14: 33: 32.9876+48: 42: 27.5872& $/$& $/$& 1.63& 1.81$\pm$0.09 & ...           & ...           & 29& 19.20$\pm$1.00 \\
...         & ...  & ...                            & ...& ...& 4.87& 0.28$\pm$0.02 & ...           & ...           & 19& 3.39$\pm$0.29 \\
J1520$+$4211& GAIA1& 15: 20: 39.7218+42: 11: 11.4524& 0.2& 0.3& 1.63& 32.50$\pm$1.63& 0.7$\pm$0.1   & $-$3.9$\pm$0.1& 29& 42.37$\pm$2.12 \\
...         & ...  & ...                            & ...& ...& 4.87& 25.96$\pm$1.30& 1.0$\pm$0.1   & $-$1.0$\pm$0.1& 13& 28.90$\pm$1.45 \\
...         & GAIA2& 15: 20: 39.6527+42: 11: 10.6437& 0.3& 0.3& 1.63& $\le$0.11    & $/$           & $/$           & 22& $/$ \\
...         & ...  & ...                            & ...& ...& 4.87& $\le$0.07    & ...           & ...           & 13& ... \\
J0928$+$5707& GAIA1& 09: 28: 53.5310+57: 07: 35.9477& 0.5& 0.4& 1.63& 3.97$\pm$0.20 & $-$0.1$\pm$0.2& 2.3$\pm$0.2   & 24& 4.09$\pm$0.21 \\
...         & ...  & ...                            & ...& ...& 4.87& 1.99$\pm$0.10 & $-$0.3$\pm$0.2& 0.9$\pm$0.2   & 18& 2.34$\pm$0.12 \\
...         & GAIA2& 09: 28: 53.4853+57: 07: 34.9952& 0.3& 0.2& 1.63& $\le$0.12    & $/$           & $/$           & 24& $/$ \\
...         & ...  & ...                            & ...& ...& 4.87& $\le$0.09    & ...           & ...           & 18& ... \\
\hline
\end{tabular}
\\
\begin{flushleft}
\tablefoot{Note: Column 1: Target name. Column 2: Feature marks for each source. Column 3: J2000 coordinate for each \textit{Gaia} position (for EXT1, the position is from its radio peak). Columns 4 and 5: The position error of the \textit{Gaia} coordinate along RA and Dec. Column 6: Frequency of the image. Column 7: The peak intensity of the image. Columns 8 and 9: The RA and Dec offsets of the VLBI core component with respect to the \textit{Gaia} position, the errors refer to the positional uncertainties derived from the VLBI imaging and model fitting (see Sect. \ref{sec:data}). Column 10: The post-CLEAN rms noise of the image within each feature. Column 11: The total flux density of the component at the position.}
\end{flushleft}

\end{sidewaystable*}

\onecolumn
\begin{sidewaystable*}
\caption{Position and radio information of the R-SMGDs.}
\label{tab:big_table}
\centering
\begin{tabular}{ccccccccc}
\hline
SDSS name & z & S$_{\rm NVSS}$ & S$_{\rm FIRST}$ & Radio coord. & Sep                & Distance & \textit{Gaia} pos. & Note from SQUAB-II \\
(J2000)  &          &  (mJy)        & (mJy/b)        & (J2000)         & ($^{\prime\prime}$)& (kpc) & (J2000) &   \\ \hline
143333$+$484227 & 1.36 & 139.7 & 65.2 & 143332.955$+$484228.13 & 0.77 & 6.48  & 14:33:33.0316$+$48:42:27.7752 & REST \\
 & $\dots$ &  &  &  & $\dots$ & $\dots$ & 14:33:32.9630$+$48:42:27.4198 & $\dots$ \\
152039$+$421111 & 0.49 & 138.4 & 53.6 & 152039.768$+$421111.68 & 1.12 & 6.76  & 15:20:39.7218$+$42:11:11.4524 & Quasar-star \\
 & $\dots$ &  &  &  & $\dots$ & $\dots$ & 15:20:39.6527$+$42:11:10.6437 & $\dots$ \\
092853$+$570735 & 1.68 & 55.4 & 20.8 & 092853.614$+$570735.74 & 1.02 & 8.64  & 09:28:53.5310$+$57:07:35.9477 & Quasar-star \\
 & $\dots$ &  &  &  & $\dots$ & $\dots$ & 09:28:53.4853$+$57:07:34.9952 & $\dots$ \\
\hline
221130$+$291949* & 0.15 & 27.4 & NA & 221130.56$+$291949.0 & 1.16 & 3.03  & 22:11:30.3379$+$29:19:48.9084 & DQ candidate \\
 & $\dots$ &  &  &  & $\dots$ & $\dots$ & 22:11:30.2526$+$29:19:49.2264 & $\dots$ \\
141546$+$112943 & 2.55 & 7.8 & 6.5 & 141546.252$+$112943.69 & 0.68, 1.10 & 5.46, 8.84 & 14:15:46.2766$+$11:29:43.3142 & LQ \\
 & $\dots$ &  &  &  & 0.84 & 6.75  & 14:15:46.2268$+$11:29:43.1233 & $\dots$ \\
 & $\dots$ &  &  &  & $\dots$ & $\dots$ & 14:15:46.2062$+$11:29:43.9142 & $\dots$ \\
163621$+$113742 & 1.38 & 4.8 & 6.2 & 163621.184$+$113743.21 & 0.23 & 1.94  & 16:36:21.1699$+$11:37:42.6643 & Quasar-star \\
 & $\dots$ &  &  &  & $\dots$ & $\dots$ & 16:36:21.1586$+$11:37:42.5093 & $\dots$ \\
155859$+$282429 & 2.3 & 5.7 & 1.9 & 155859.147$+$282428.97 & 0.6 & 4.92  & 15:58:59.1229$+$28:24:28.7780 & Quasar-star \\
 & $\dots$ &  &  &  & $\dots$ & $\dots$ & 15:58:59.1040$+$28:24:29.3230 & $\dots$ \\
080009$+$165509 & 1.59 & 4.1 & 2.9 & 080009.971$+$165510.22 & 0.77 & 6.52  & 08:00:10.0102$+$16:55:09.1339 & LQ candidate \\
 & $\dots$ &  &  &  & $\dots$ & $\dots$ & 08:00:09.9695$+$16:55:09.6304 & $\dots$ \\
072843$+$370835 & 1.37 & 3.7 & 1.1 & 072843.019$+$370835.05 & 1.36 & 11.45  & 07:28:43.1054$+$37:08:35.7161 & Quasar-star \\
 & $\dots$ &  &  &  & $\dots$ & $\dots$ & 07:28:43.0174$+$37:08:34.8611 & $\dots$ \\
125631$+$330253 & 2.56 & 2.7 & 1.7 & 125631.345$+$330252.81 & 0.51 & 4.09  & 12:56:31.3502$+$33:02:52.8828 & Quasar-star \\
 & $\dots$ &  &  &  & $\dots$ & $\dots$ & 12:56:31.3490$+$33:02:53.3878 & $\dots$ \\
162501$+$430931 & 0.71 & 2.5 & 2.6 & 162501.973$+$430931.50 & 0.57 & 4.10  & 16:25:01.9904$+$43:09:31.3940 & LQ candidate \\
 & $\dots$ &  &  &  & $\dots$ & $\dots$ & 16:25:01.9651$+$43:09:31.8958 & $\dots$ \\
172308$+$524455 & 1.66 & NA & 2.1 & 172308.051$+$524456.37 & 1.1 & 9.32  & 17:23:08.1644$+$52:44:55.1994 & Quasar-star \\
 & $\dots$ &  &  &  & $\dots$ & $\dots$ & 17:23:08.0892$+$52:44:56.0643 & $\dots$ \\
080037$+$461257 & 1.82 & NA & 2 & 080037.608$+$461257.80 & 0.88 & 7.43  & 08:00:37.6612$+$46:12:57.9840 & Quasar-star \\
 & $\dots$ &  &  &  & $\dots$ & $\dots$ & 08:00:37.5793$+$46:12:57.7405 & $\dots$ \\
163348$+$313411 & 1.52 & NA & 1.8 & 163348.924$+$313411.66 & 0.66 & 5.59  & 16:33:49.0184$+$31:34:11.6082 & LQ \\
 & $\dots$ &  &  &  & $\dots$ & $\dots$ & 16:33:48.9775$+$31:34:12.0123 & $\dots$ \\
000710$+$005329 & 0.32 & NA & 1.7 & 000710.011$+$005328.56 & 0.79 & 3.68  & 00:07:10.0179$+$00:53:28.9972 & REST \\
 & $\dots$ &  &  &  & $\dots$ & $\dots$ & 00:07:09.9728$+$00:53:29.4076 & $\dots$ \\
095122$+$263513 & 1.25 & NA & 1.5 & 095122.539$+$263514.11 & 1.1 & 9.17  & 09:51:22.6363$+$26:35:13.4022 & LQ candidate \\
 & $\dots$ &  &  &  & $\dots$ & $\dots$ & 09:51:22.5695$+$26:35:14.0383 & $\dots$ \\
221217$+$035040 & 0.21 & NA & 1.1 & 221217.090$+$035041.30 & 0.93 & 3.19  & 22:12:17.1165$+$03:50:40.5459 & LQ candidate \\
 & $\dots$ &  &  &  & $\dots$ & $\dots$ & 22:12:17.0789$+$03:50:41.2862 & $\dots$ \\
091301$+$525928 & 1.38 & NA & 1.1 & 091301.064$+$525930.05 & 1.1 & 9.27  & 09:13:01.1117$+$52:59:28.4941 & LQ \\
 & $\dots$ &  &  &  & $\dots$ & $\dots$ & 09:13:01.0025$+$52:59:28.9907 & $\dots$ \\
 \hline
\end{tabular}
\\
\begin{flushleft}
\tablefoot{Column 1: J2000 name from the SDSS. Column 2: Redshifts. Column 3: Flux density from NVSS. Column 4: Peak flux density form FIRST. Column 5: Coordinates from FIRST or NVSS (The one nearest to the optical core are chosen). Column 6: Angular separation of the two \textit{Gaia} positions (the later one(s) with respect to the current one). Column 7: Physical distance of the two positions in kpc. Column 8: The high accuracy positions from \textit{Gaia} EDR3. Column 9: The identification note from the SQUAB-II paper.\\ * This source is not in the FIRST survey footprint.}
\end{flushleft}
\end{sidewaystable*}

\twocolumn
\onecolumn
\section{The VLBI big images of the target sources at 1.6 GHz}
\label{app:wideim}
\begin{figure*}[h!]
    \centering
    \includegraphics[width=0.9\linewidth]{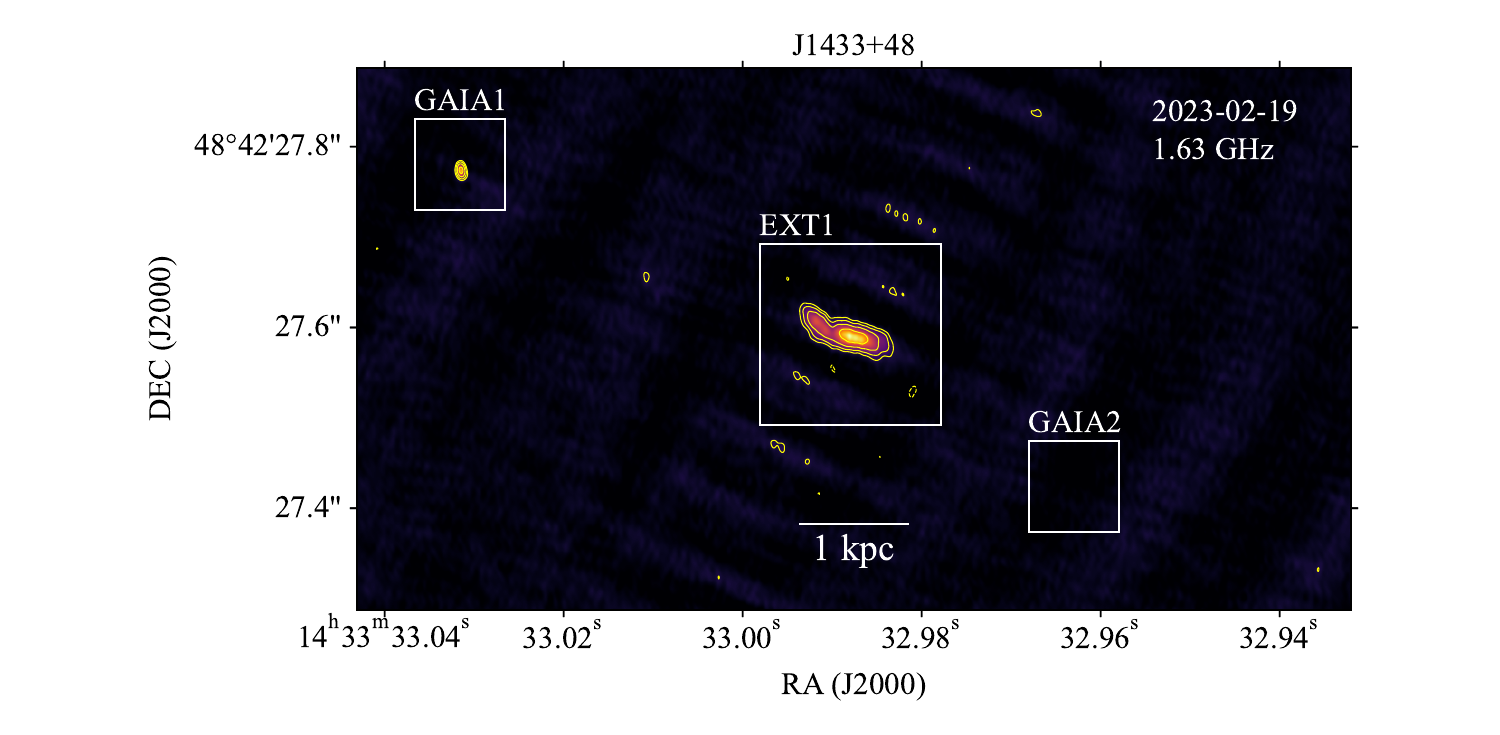} 
    \caption{\label{fig:J1433_big}Naturally weighted wide view VLBI image of J1433$+$4842 at 1.63 GHz. It includes the first \textit{Gaia} position (GAIA1) the second \textit{Gaia} position (GAIA2) and the extended emission that lies between the two positions (EXT1). To note that the shape of EXT1 in the picture is not exact the same with the zooming-in images showing in Fig. \ref{fig:J1433_small} due to the bandwidth smearing effect caused by the large distance between the source and the image phase center (I used the GAIA1 position as the big image's phase center). The lowest contours represent 4$\sigma$ ($\sigma$ the post-CLEAN background noise) and contour levels increase by a factor of 2.}
\end{figure*}

\begin{figure*}[h!]
    \centering
    \includegraphics[width=12cm]{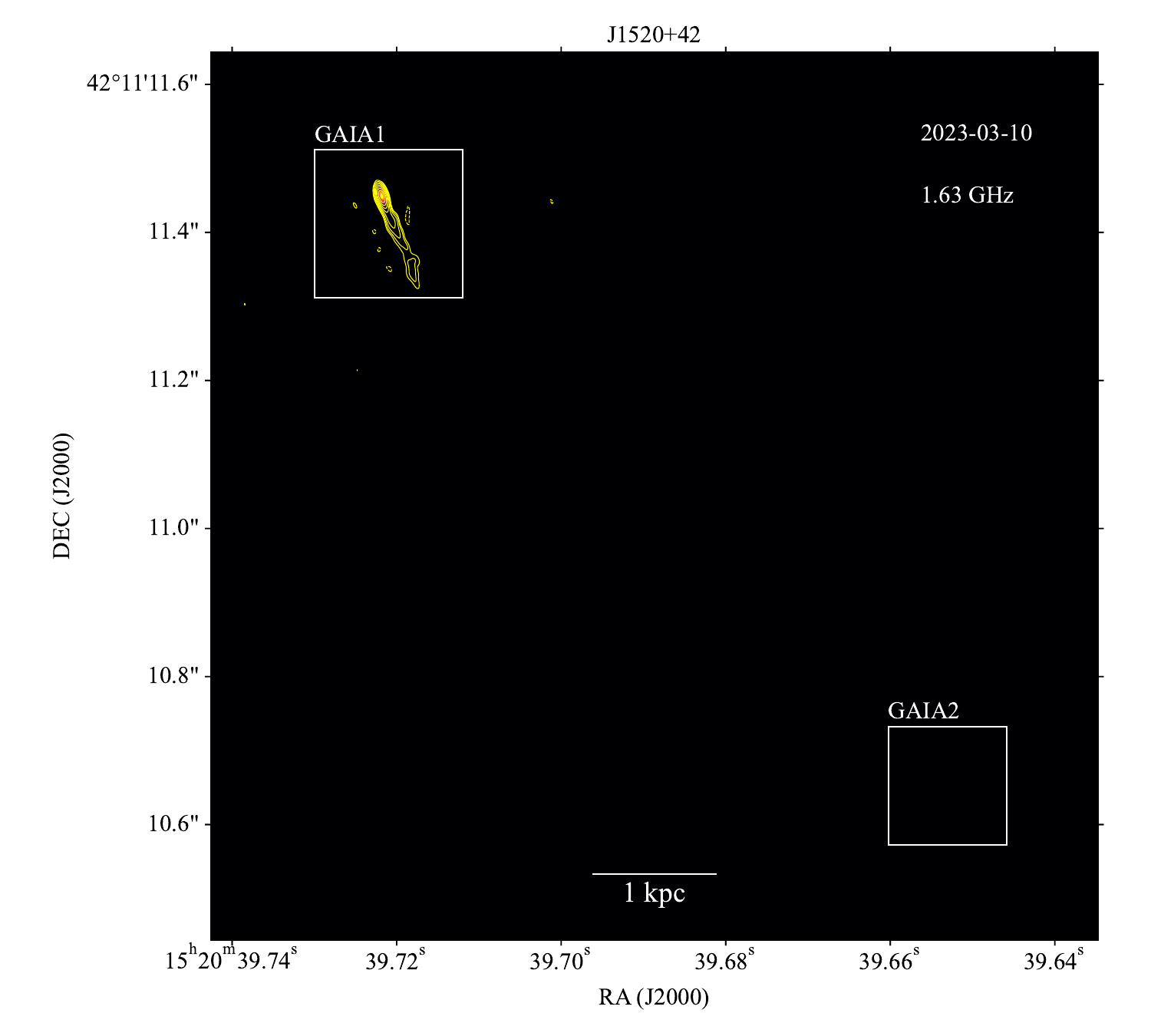} 
    \caption{\label{fig:J1520_big}Naturally weighted wide view VLBI image of J1520$+$4211 at 1.63 GHz showing both GAIA1 and GAIA2 positions. The lowest contours represent 4$\sigma$ and contour levels increase by a factor of 2. The single-source zooming-in image of each position are shown in Fig. \ref{fig:J1520_small}.}
\end{figure*}

\begin{figure*}[h!]
    \centering
    \includegraphics[width=12cm]{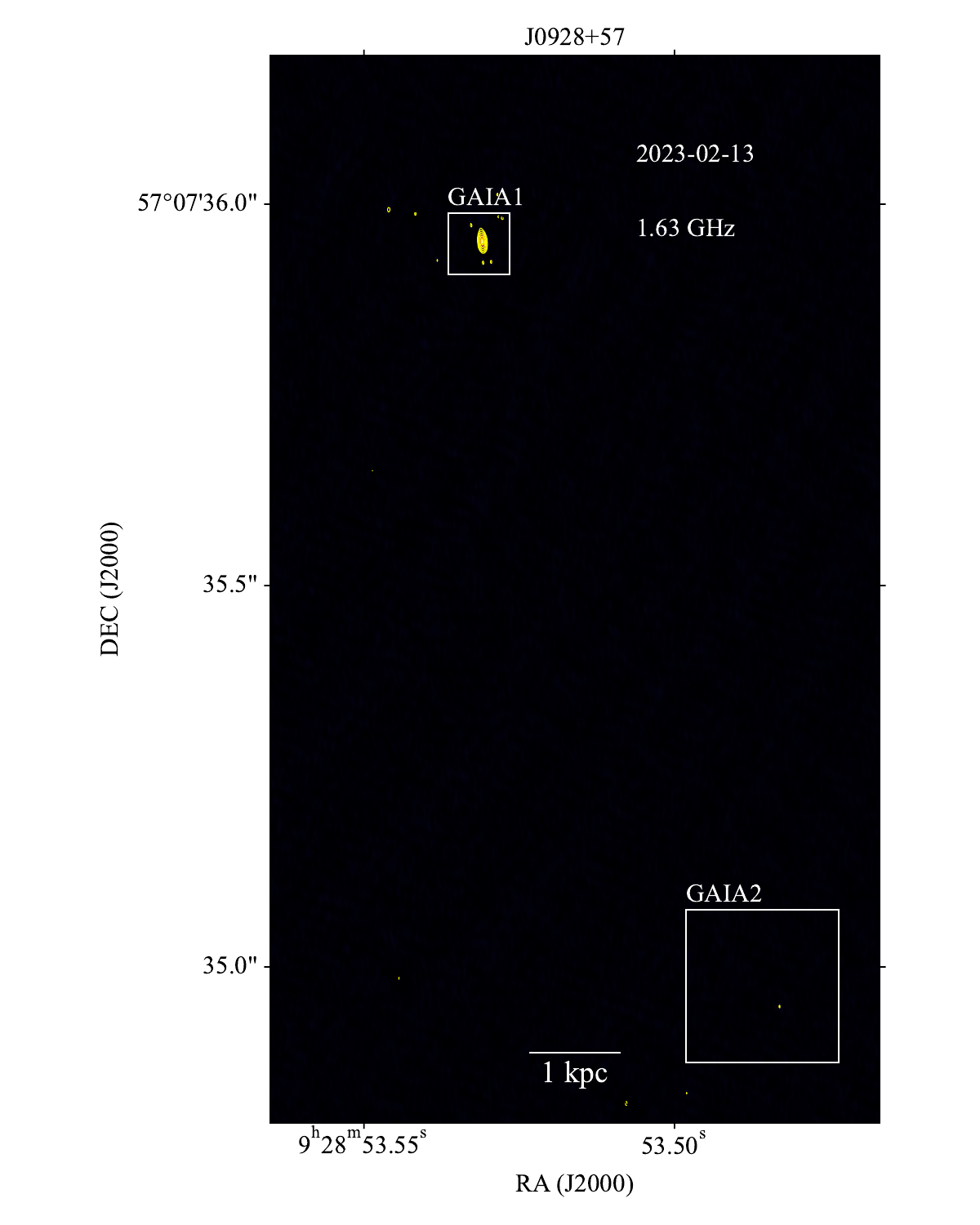} 
    \caption{\label{fig:J0928_big}Naturally weighted wide view VLBI image of J0928$+$5707 at 1.63 GHz showing both GAIA1 and GAIA2 positions. The lowest contours represent 4$\sigma$ and contour levels increase by a factor of 2. The single-source zooming-in image of each position are shown in Fig. \ref{fig:J0928_small}.}
\end{figure*}

\twocolumn
\onecolumn

\section{VLBI zoomed-in images of the target sources at 1.6 GHz and 4.9 GHz}
\label{app:smallim}
\begin{figure*}[h!]
    \centering
    \includegraphics[width=0.35\linewidth]{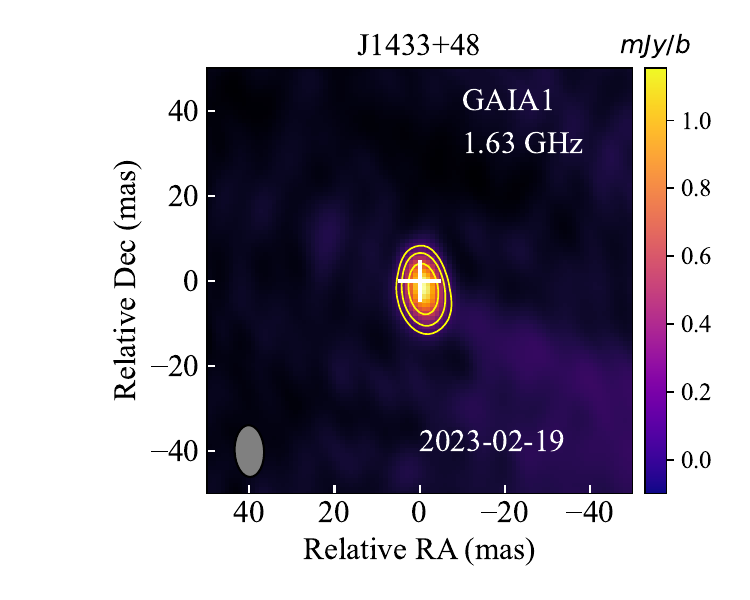}
    \includegraphics[width=0.35\linewidth]{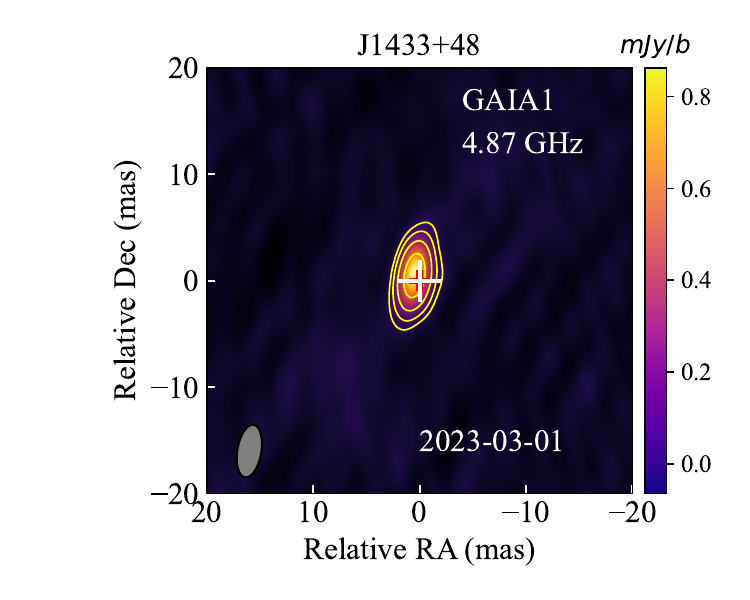} \\
    \includegraphics[width=0.35\linewidth]{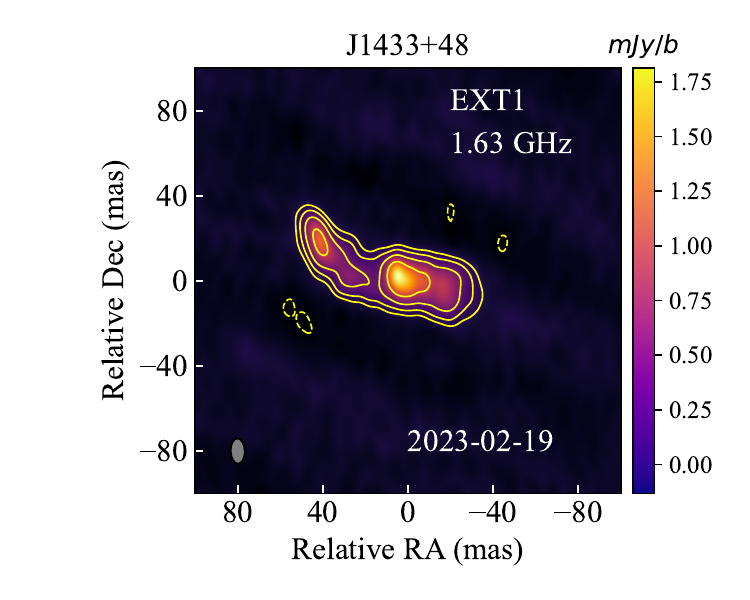}
    \includegraphics[width=0.35\linewidth]{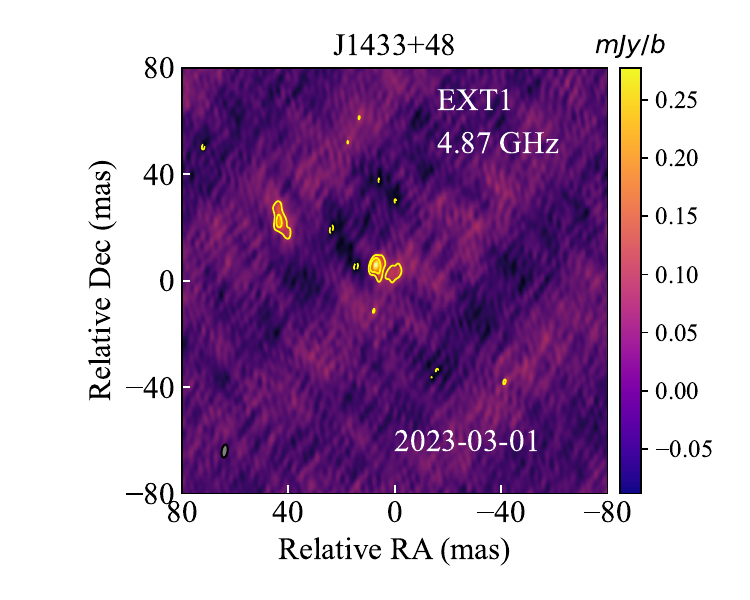} \\
    \includegraphics[width=0.35\linewidth]{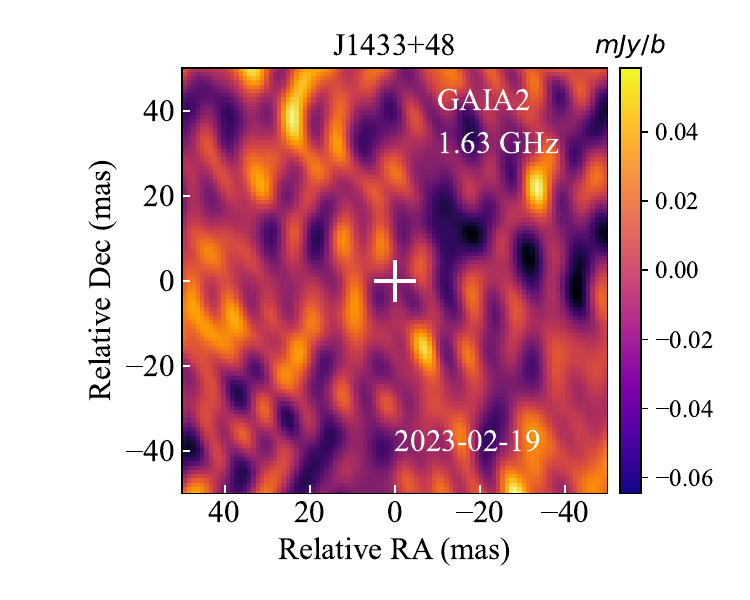}
    \includegraphics[width=0.35\linewidth]{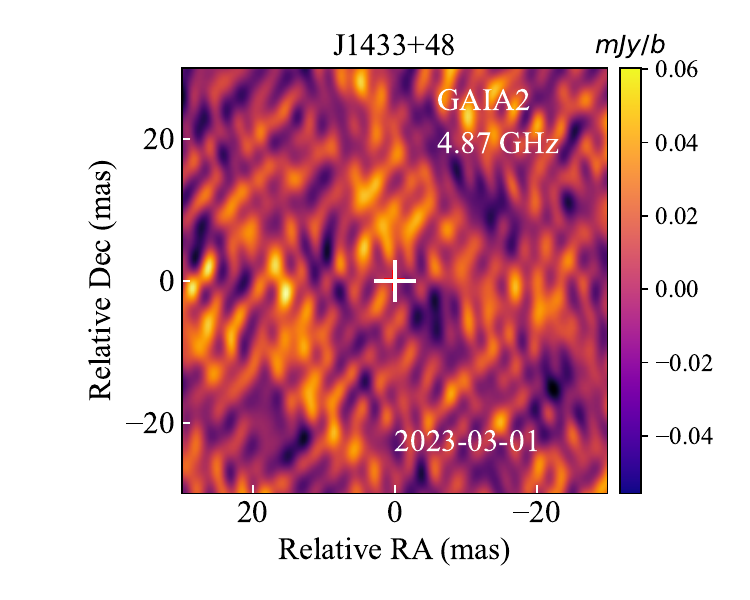} 
    \caption{Naturally weighted VLBI single-source images of each target position for J1433$+$4842 at 1.7 and 4.9 GHz. The white crosses in the GAIA1 and GAIA2 images mark the corresponding \textit{Gaia} positions in each source position. For each CLEANed image, the synthesized beam shape are shown at the bottom left corner. The lowest contours represent 4$\sigma$ and contour levels increase by a factor of 2. For the GAIA2 images, no significant signal can be detected above 5$\sigma$. For this source in the image, 1 mas represents 8.413 $\rm pc$.\label{fig:J1433_small}}
\end{figure*}

\begin{figure*}
    \centering
    \includegraphics[width=0.35\linewidth]{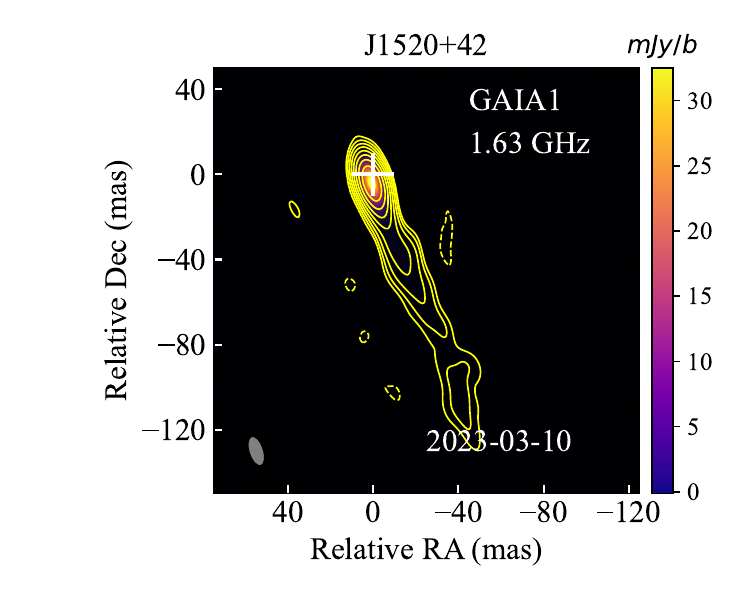}
    \includegraphics[width=0.35\linewidth]{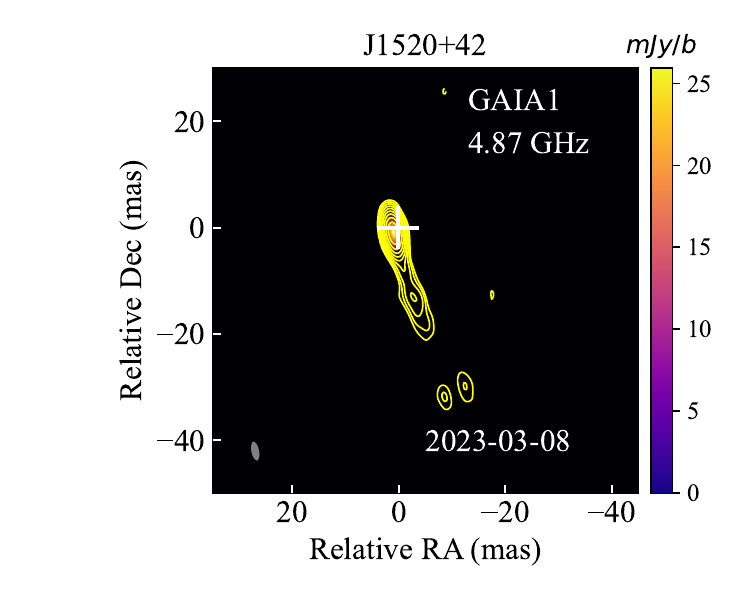} \\
    \includegraphics[width=0.35\linewidth]{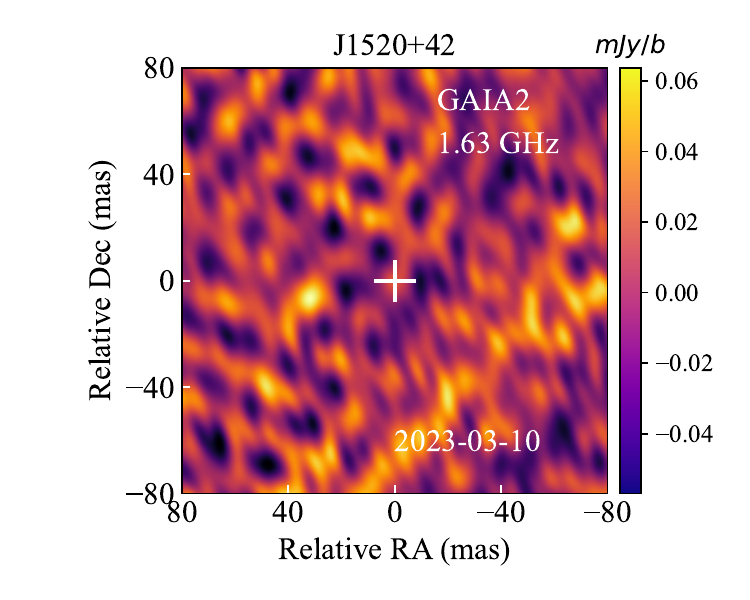}
    \includegraphics[width=0.35\linewidth]{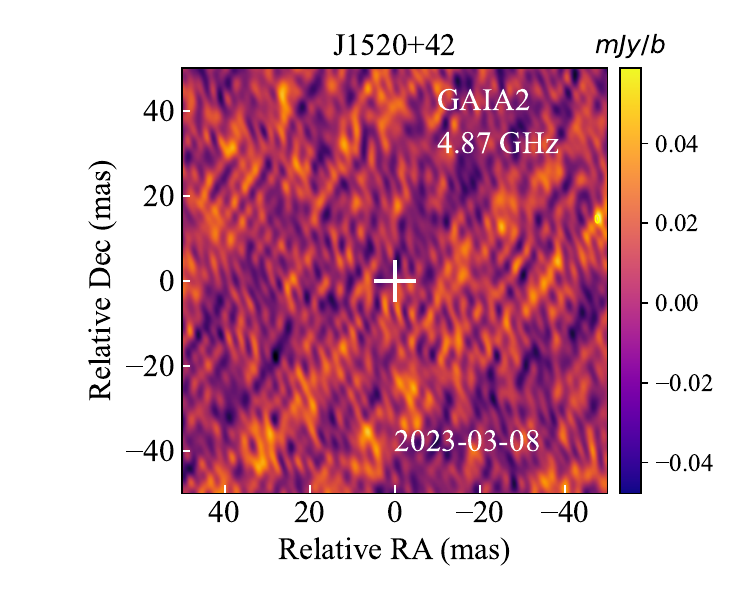} 
    \caption{\label{fig:J1520_small}Naturally weighted single-source VLBI images of J1520$+$4211 at 1.7 and 4.9 GHz. The imaging notes are the same as in Fig. \ref{fig:J1433_small}. The scale factor of this source is 6.038 $pc$/mas}
\end{figure*}
\begin{figure*}
    \centering
    \includegraphics[width=0.35\linewidth]{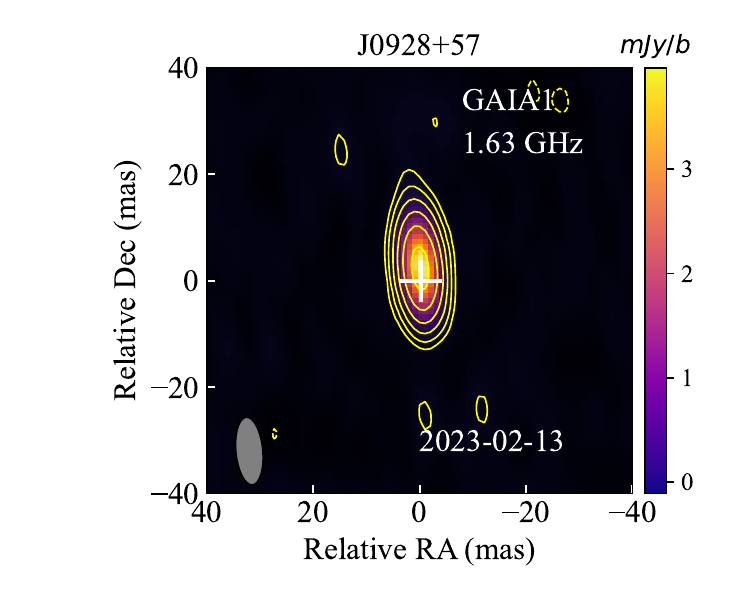}
    \includegraphics[width=0.35\linewidth]{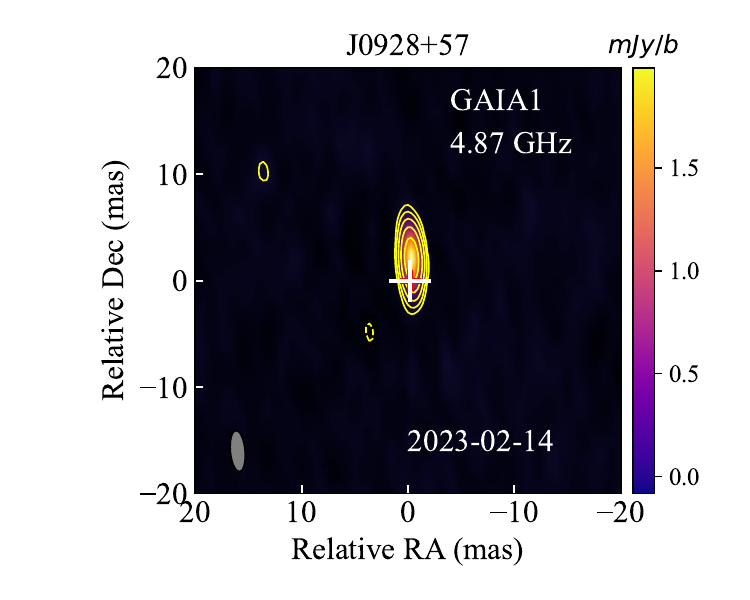} \\
    \includegraphics[width=0.35\linewidth]{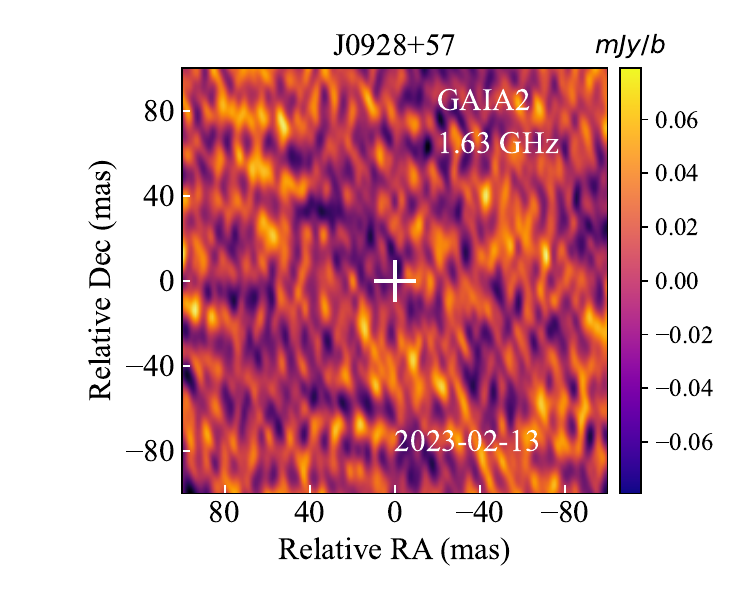}
    \includegraphics[width=0.35\linewidth]{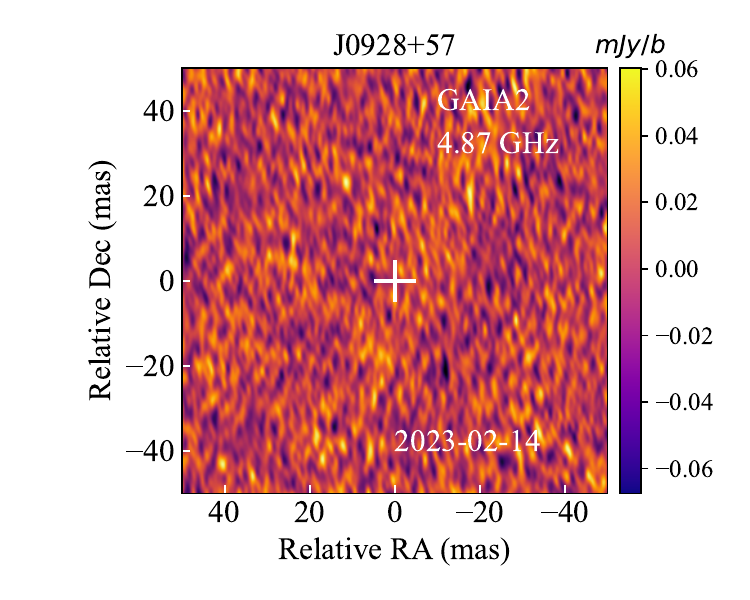} \\
    \caption{\label{fig:J0928_small}Naturally weighted single-source VLBI images of J0928$+$5707 at 1.7 and 4.9 GHz. The imaging notes are the same as in Fig. \ref{fig:J1433_small}. 1 mas represents 8.467 in the above images.}
\end{figure*}

\twocolumn
\onecolumn
\section{Additional figures}
\begin{figure*}[ht!]
    \centering
    \includegraphics[width=0.33\linewidth]{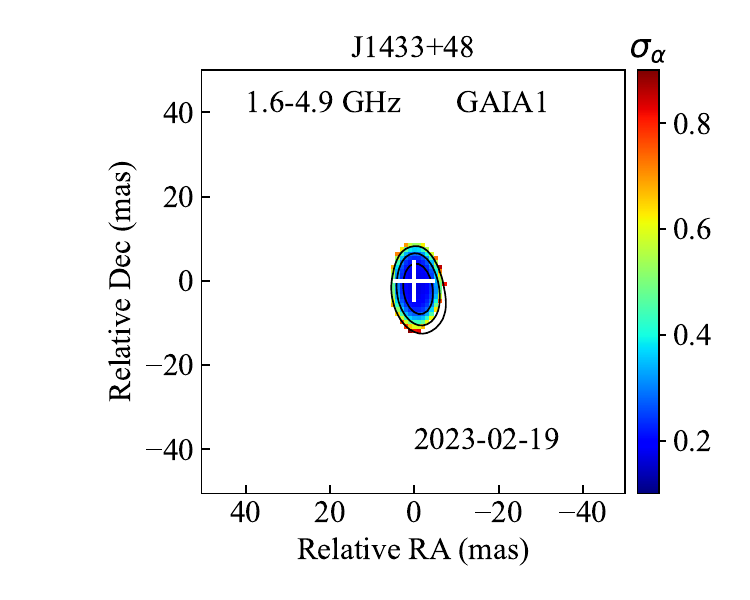} 
    \includegraphics[width=0.33\linewidth]{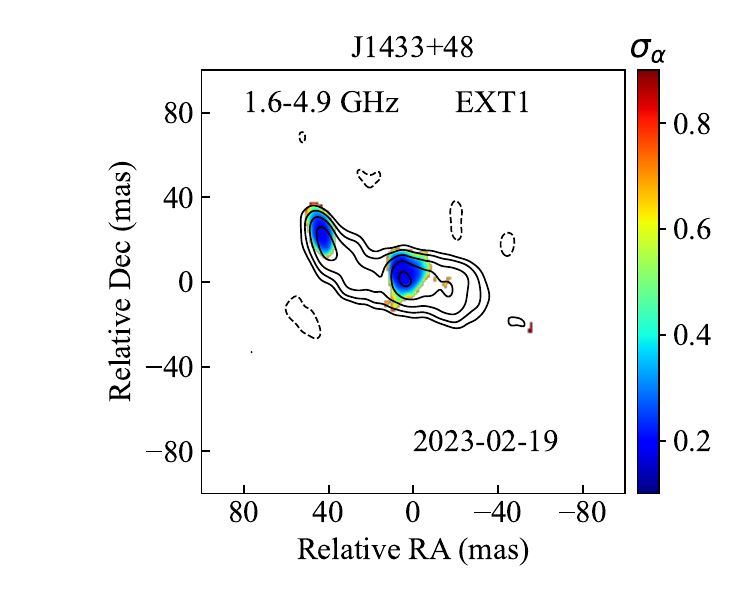} \\
    \includegraphics[width=0.33\linewidth]{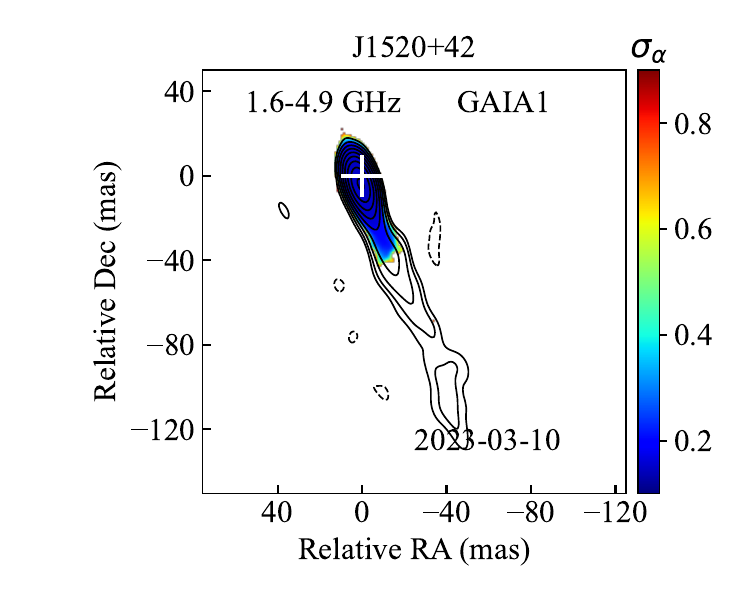} 
    \includegraphics[width=0.33\linewidth]{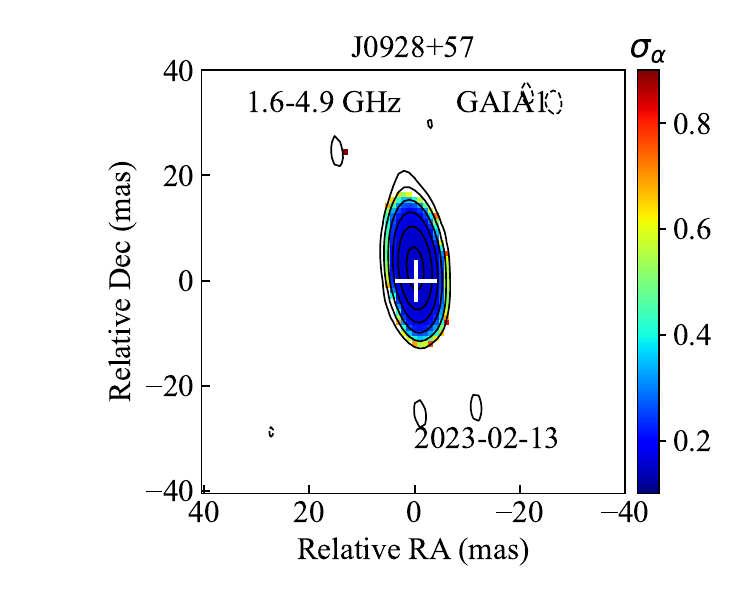} 
    \caption{Spectral index error distributions corresponding to the spectral index maps for Fig. \ref{fig:spix}.\label{fig:spixerr}}
\end{figure*}

\begin{figure*}[hb!]
    \centering
    \includegraphics[width=0.31\linewidth]{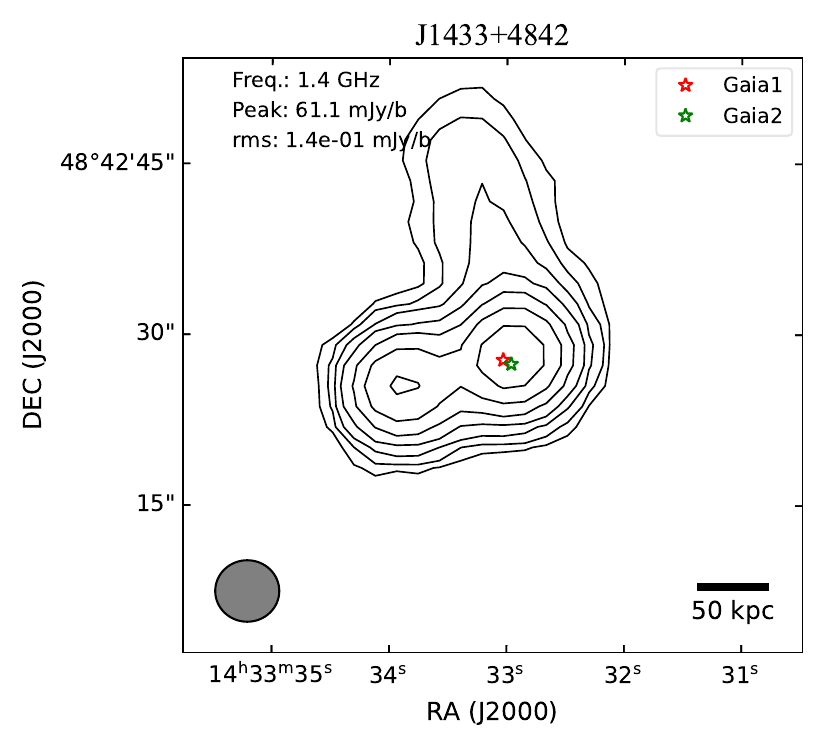} 
    \includegraphics[width=0.31\linewidth]{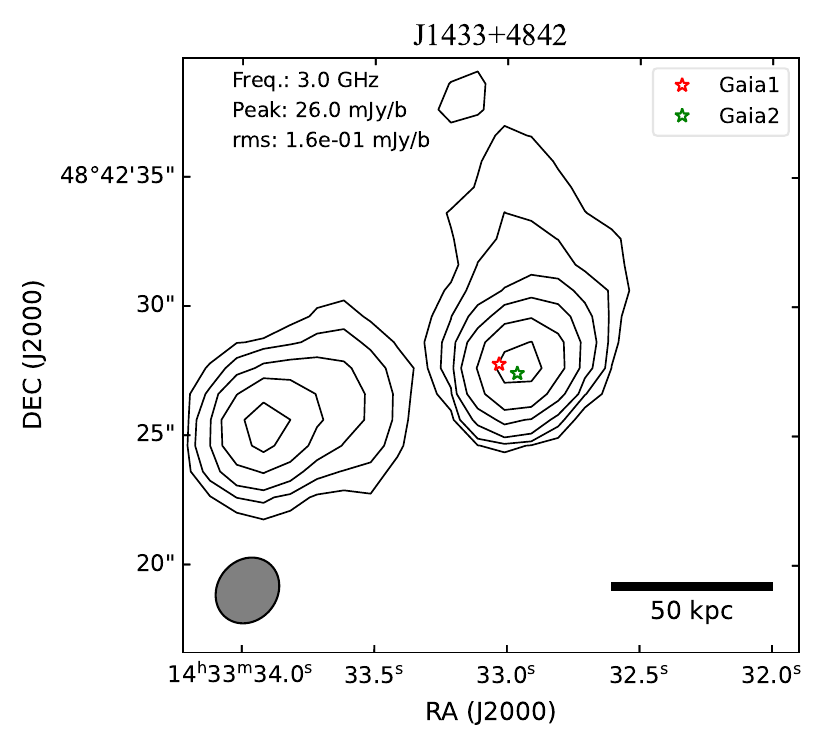} \\
    \includegraphics[width=0.31\linewidth]{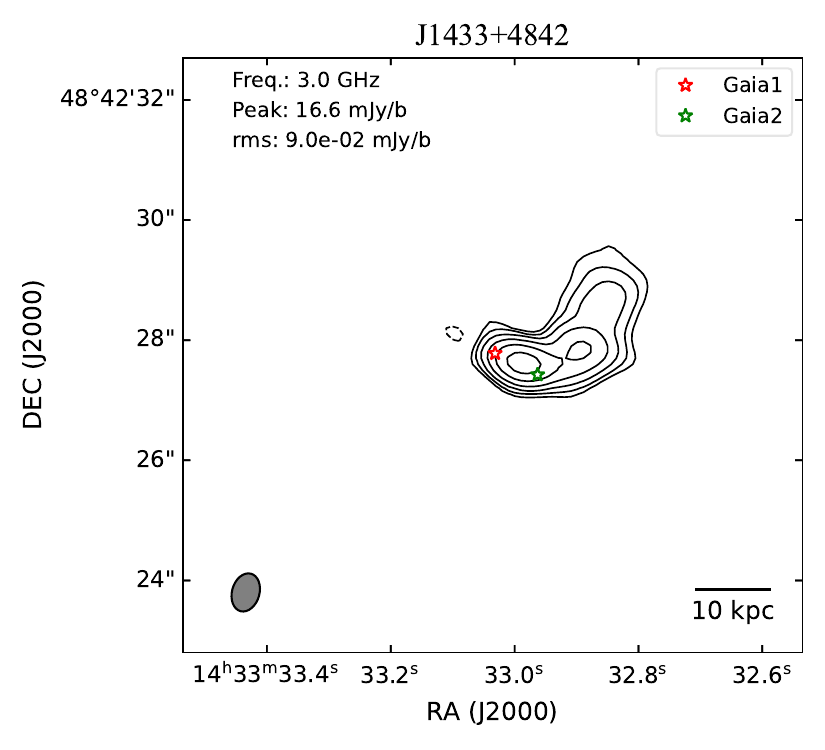}
    \includegraphics[width=0.31\linewidth]{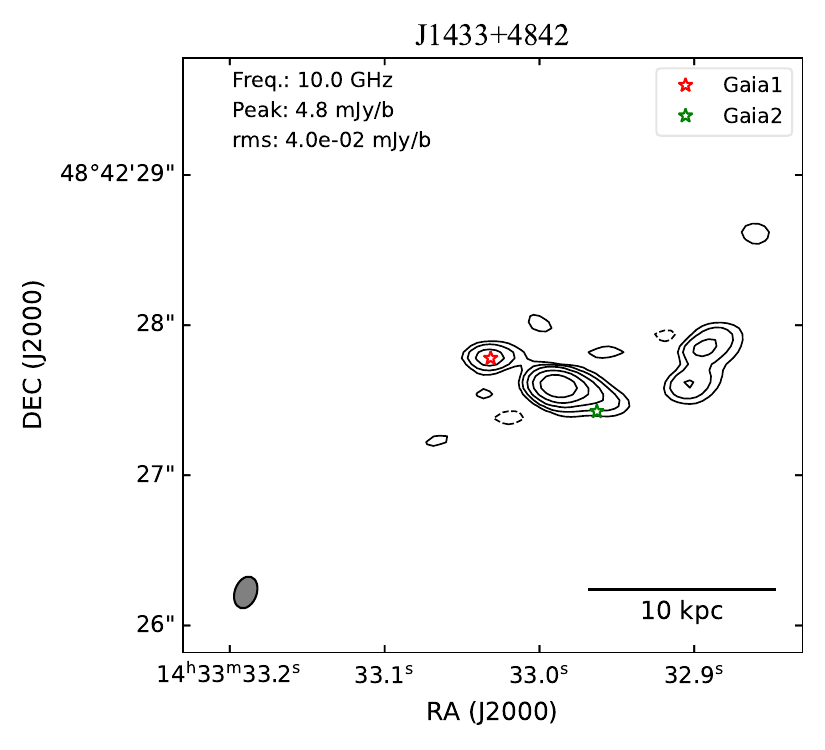}
    \caption{Naturally weighted large scale radio structure for J1433$+$4842. The upper two panels are derived from the NRAO FIRST (1.4 GHz) and VLASS (3 GHz) survey. The bottom two are from the VLA observations from S24 at 3 and 10 GHz. The stars marked the \textit{Gaia} positions. The lowest contours are 4 times of the background noise and contours increase by a factor of two. The beams are shown at the bottom left corner of each image.\label{fig:target1_vla}}
\end{figure*}

\end{appendix}
\end{document}